\documentstyle[12pt,aaspp4,flushrt,eqsecnum,epsfig]{article}
\newcommand{\clb}{{\mathcal{B}}}
\newcommand{\clc}{{\mathcal{C}}}

\newcommand{\cle}{{\mathcal{E}}}
\newcommand{\clh}{{\mathcal{H}}}
\newcommand{\cll}{{\mathcal{L}}}
\newcommand{\clr}{{\mathcal{R}}}
\newcommand{\clt}{{\mathcal{T}}}
\newcommand{\clw}{{\mathcal{W}}}
\newcommand{\clx}{{\mathcal{X}}}
\newcommand{\clz}{{\mathcal{Z}}}
\newcommand{\mpc}{\mbox{\rm \,Mpc}}
\newcommand{\kmsmpc}{\mbox{$\mathrm{kms^{-1}Mpc^{-1}}$}}
\newcommand{\fgam}{f^{(\gamma)}}
\newcommand{\Fgaml}{F^{(l)}}
\newcommand{\Fgamlin}[1]{F^{(#1)}}
\newcommand{\Jgamlin}[1]{J^{(#1)}}
\newcommand{\Jgamlink}[1]{J^{(#1)}_{k}}
\newcommand{\Jgaml}{J^{(l)}}
\newcommand{\Jgamlk}{\Jgaml_{k}}
\newcommand{\Jgamlkdt}{\dot{J}^{(l)}_{k}}
\newcommand{\Tgamlk}{T^{(l)}(k)}
\newcommand{\Gnulin}[1]{G^{(#1)}}
\newcommand{\Gnul}{G^{(l)}}
\newcommand{\Gnulk}{\Gnul_{k}}
\newcommand{\Gnulkdt}{\dot{G}^{(l)}_{k}}
\newcommand{\Gnulink}[1]{G^{(#1)}_{k}}
\newcommand{\nelec}{n_{e}}
\newcommand{\csound}{c_{s}}
\newcommand{\qk}{Q^{(k)}}
\newcommand{\qkdt}{\dot{Q}^{(k)}}

\newcommand{\qkstar}{Q^{(k)\ast}}
\newcommand{\qka}{Q^{(k)}_{a}}

\newcommand{\qkab}{Q^{(k)}_{ab}}

\newcommand{\sgrad}{{}^{(3)}\nabla}
\newcommand{\tfrac}[2]{\case{#1}{#2}}
\newcommand{\half}{\case{1}{2}}
\newcommand{\et}[1]{e^{\mbox{\footnotesize $#1$}}}
\newcommand{\ord}{\mbox{O}}
\newcommand{\eqref}[1]{(\ref{#1})}
\newcommand{\bk}{\mbox{\boldmath $k$}}
\newcommand{\bx}{\mbox{\boldmath $x$}}
\newcommand{\snthet}{\sin\!\theta}
\newcommand{\csthet}{\cos\!\theta}
\newcommand{\snphi}{\sin\!\phi}
\newcommand{\csphi}{\cos\!\phi}
\lefthead{Challinor \&\ Lasenby}
\righthead{Covariant and gauge-invariant CMB anisotropies}
 
\begin{document}
 
\title{COSMIC MICROWAVE BACKGROUND ANISOTROPIES IN THE CDM MODEL: A
COVARIANT AND GAUGE-INVARIANT APPROACH}
\author{Anthony Challinor\footnote{A.D.Challinor@mrao.cam.ac.uk} \&\
Anthony Lasenby\footnote{A.N.Lasenby@mrao.cam.ac.uk}}
\affil{Astrophysics Group, Cavendish Laboratory, Madingley Road, Cambridge CB3 0HE, UK.}
 
\begin{abstract}

We present a fully covariant and gauge-invariant calculation of the evolution
of anisotropies in the Cosmic Microwave Background (CMB) radiation.
We use the physically appealing covariant approach to cosmological
perturbations, which ensures that all variables are gauge-invariant
and have a clear physical interpretation. We derive the complete set of
frame-independent linearised equations describing the (Boltzmann) evolution of
anisotropy and inhomogeneity in an almost Friedmann-Robertson-Walker (FRW)
Cold Dark Matter (CDM) universe.
These equations include the contributions of scalar, vector and tensor
modes in a unified manner. Frame-independent equations for scalar and tensor
perturbations, which are valid for any value of the background curvature,
are obtained straightforwardly from the complete set of equations.
We discuss the scalar equations in detail, including the integral solution and
relation with the line of sight approach, analytic solutions in the
early radiation dominated era, and the numerical solution in the standard CDM
model. Our results confirm those obtained by
other groups, who have worked carefully with non-covariant methods in
specific gauges, but are derived here in a completely transparent fashion.

\end{abstract}

\keywords{cosmic microwave background --- cosmology: theory --- gravitation
--- large-scale structure of universe}

\section{Introduction}

The cosmic microwave background radiation (CMB) occupies a central role in
modern cosmology. It provides us with a unique record of conditions along our
past lightcone back to the epoch of decoupling (last scattering), when the
optical depth to Thomson scattering rises suddenly due to Hydrogen
recombination. Accurate observations of the CMB anisotropy should allow us to
distinguish between models of structure formation and, in the case of
non-seeded models, to infer the spectrum of initial perturbations in the
early universe. Essential to this programme is the accurate and reliable
calculation of the anisotropy predicted in viable cosmological models.

Such calculations have a long history, beginning with Sachs \&\
Wolfe (1967) who investigated the anisotropy on large scales
$(\gtrsim 1^{\circ})$ by calculating the redshift back to last scattering
along null geodesics in a perturbed universe. On smaller angular scales
one must address the detailed local processes occurring in the electron/baryon
plasma prior to recombination, and the effects of non-instantaneous last
scattering. These processes, which give rise to a wealth of structure in the
CMB power spectrum on intermediate scales and damping on small
scales (see, for example, Silk (1967, 1968)),
are best addressed by following the photon distribution function directly
from an early epoch in the history of the universe to the current point of
observation. This requires a numerical integration of the Boltzmann equation,
and has been carried out by many groups,
of which Peebles \&\ Yu (1970), Bond \&\ Efstathiou (1984, 1987), Hu \&\
Sugiyama (1995), Ma \&\ Bertshcinger (1995), Seljak \&\ Zaldarriaga (1996)
is a representative sample.

The calculation of CMB anisotropies is simple in principle, but in reality
is plagued with subtle gauge issues (Stoeger, Ellis, \&\ Schmidt 1991;
Stoeger et al.\ 1995; Challinor \&\ Lasenby 1998).
These problems arise because of the gauge-freedom in specifying a map
$\Phi$ between the real universe (denoted by $S$)
and the unperturbed background model(denoted by $\bar{S}$)~\cite{ellis89a},
which is usually taken
to be a Friedmann-Robertson-Walker (FRW) universe. The map $\Phi$
identifies points
in the real universe with points in the background model, thus defining the
perturbation in any quantity of interest. For example, for the density $\rho$
as measured by some physically defined observer, the perturbation at
$x\in S$ is defined to be $\delta \rho(x)\equiv \rho(x) - \bar{\rho}(\bar{x})$,
where $\bar{\rho}$ is the equivalent density in the background model, and
$x$ maps to $\bar{x}$ under $\Phi$. The map $\Phi$ is usually (partially)
specified by imposing coordinate conditions in $S$ and $\bar{S}$. Any
residual freedom in the map $\Phi$ after the imposition of the coordinate
conditions (gauge-fixing) gives rise to the following gauge problems:
(i) the map cannot be reconstructed from observations in $S$ alone, so that
quantities such as the density perturbation, which depend on the specific
map $\Phi$, are necessarily not observable; (ii) if the residual
gauge freedom allows points in $\bar{S}$ to be mapped to physically
inequivalent points in $\bar{S}$ in the limit that $S=\bar{S}$, then
unphysical gauge mode solutions to the linearised perturbation equations
will exist.

There are several ways to deal with the gauge problems described above.
In the earliest approach~\cite{lifshitz46}, one retains the residual gauge
freedom (in the synchronous gauge) but keeps track of it so that gauge
mode solutions can be eliminated. Furthermore, the final results of such
a calculation must be expressed in terms of the physically relevant,
gauge-invariant quantities. Although there is nothing fundamentally wrong
with this approach if carried out correctly, it suffers from a long history
littered with confusion and errors. The need to express results in terms of
gauge-invariant variables suggests that it might be beneficial to employ
such variables all along as the dynamical degrees of freedom in the
calculation. A further advantage of such an approach is that gauge modes
are automatically eliminated from the perturbation equations when expressed
in terms of gauge-invariant variables.
This is the approach adopted by Bardeen (1980), who showed
how to construct gauge-invariant variables for scalar, vector and tensor
modes in linearised perturbation theory, by taking suitable linear combinations
of the gauge-dependent perturbations in the metric and matter variables.
This approach has been used in several calculations of the CMB
anisotropy (see, for example, Abbott \&\ Schaefer (1986) and
Panek (1986)). However, the Bardeen variables are not entirely
satisfactory. The approach is inherently linear, so that the variables are
only defined for small departures from FRW symmetry. Furthermore, the
approach assumes a non-local decomposition of the perturbations into
scalar, vector and tensor modes at the outset, each of which is then treated
independently. As a result, the Bardeen variables are only gauge-invariant for
the restricted class of gauge-transformations that respect the scalar,
vector and tensor splitting. Finally, although the Bardeen variables are
gauge-invariant, they are not physically transparent, in that, in
a general gauge, they do not characterise the perturbations in a manner that
is amenable to simple physical interpretation.

An alternative scheme for the gauge-invariant treatment of cosmological
perturbations was given by Ellis \&\ Bruni (1989) (see also,
Ellis, Hwang, \&\ Bruni (1989)) who built upon earlier work
by Hawking (1966). In this approach, which is derived from the
covariant approach to cosmology/hydrodynamics of Ehlers and Ellis
(Ehlers 1993; Ellis 1971), the perturbations are described by gauge-invariant
variables that are covariantly defined in the real universe. This ensures
that the variables have simple physical interpretations in terms of
the inhomogeneity and anisotropy of the universe.
Since the definition of the
covariant variables does not assume any linearisation, exact equations can
be found for their evolution, which can then be linearised around the
chosen background model. Furthermore, the covariant approach does
not employ the non-local decomposition into scalar, vector or tensor modes,
at a fundamental level.
If required, the decomposition can be performed at a late stage
in the calculation to aid solving the equations.
Even if one denies that working with gauge-invariant
variables is a significant advantage, the key advantage of the covariant
approach, however, is that one is able to work exclusively with physically
relevant quantities, satisfying equations that make manifest their physical
consequences.

The covariant and gauge-invariant approach has already been applied to
the line of sight calculation of CMB anisotropies under the
instantaneous recombination approximation (Dunsby 1997; Challinor
\&\ Lasenby 1998), and has been used to obtain model-independent
limits on the inhomogeneity and anisotropy from measurements
of the CMB anisotropy on large scales~\cite{maartens95}.
In this paper, we extend the methodology developed in these earlier papers,
to give a full kinetic theory calculation of CMB anisotropies valid on
all angular scales. Our motivation for reconsidering this problem is two-fold.
Firstly, it is our belief that the covariant and gauge-invariant description
of cosmological perturbations provides a powerful set of tools for the
formulation of the basic perturbation equations, and their subsequent
interpretation, which are superior to the techniques usually employed in
such calculations for the reasons discussed above. In particular,
by applying covariant methods for the problem of the generation of CMB
anisotropies, we can expect the same advantages of physical clarity and
unification that have already been demonstrated in other
areas, (Ellis et al.\ 1989; Bruni, Ellis, \&\ Dunsby 1992; Dunsby, Bruni, \&\
Ellis 1992; Dunsby, Bassett, \&\ Ellis 1996; Tsagas \&\ Barrow 1997).
The approach described here brings the underlying physics to the fore,
and can only help to consolidate our rapidly growing understanding of the
physics of CMB anisotropies. Furthermore, although we only consider the
linearised calculation here, the extension of these methods
to the full non-linear case is quite straightforward
(Maartens, Gebbie, \&\ Ellis 1998).
Our second motivation is to perform an independent verification of the
results of other groups (for example, Ma \&\ Bertschinger (1995)), with
a methodology that is free from any of the gauge ambiguities that have
caused problems and confusion in the past. Given the potential impact
on cosmology of the next generation of CMB data, we believe that the
above comments provide ample justification for reconsidering this problem.

For definiteness we consider the cold dark matter
(CDM) model, although the methods we describe are straightforward to extend
to other models. We have endeavoured to make this paper reasonably
self-contained, so we begin with a brief overview of the covariant approach to
cosmology and define the key variables we use to characterise
the perturbations in Section~\ref{sec_cov}. We then go on to present a
complete set of frame-independent equations describing the evolution of the
matter components and radiation in Section~\ref{sec_eqs} in an almost-FRW
universe (with arbitrary spatial curvature). These equations, which employ
only covariantly defined, gauge-invariant variables, are independent of
any harmonic analysis; they describe scalar, vector and tensor
perturbations in a unified manner. Many of the equations have simple
Newtonian analogues, and their physical consequences are far more
transparent than the equations that underlie the metric-based approaches.
Equations pertinent to scalar modes, see Section~\ref{sec_scal}, and tensor
modes, see Section~\ref{sec_tens}, can be obtained from the
full set of equations with very little effort, and are useful at this late
stage in the calculation as an aid to solving the linearised equations.
A significant feature of this approach is that a covariant angular
decomposition of the distribution functions is made early on in the
calculation, before any splitting into scalar, vector and tensor modes.
This allows scalar, vector and tensor modes to be treated in a more
unified manner. In particular, the azimuthal
dependence of the moments of the distribution functions does not have to
be put in by hand (after inspection of the azimuthal dependence of the
other terms in the Boltzmann equation), as happens in most metric-based
calculations. This is particularly significant for tensor modes where
the required azimuthal dependence is non-trivial and is different for the
two polarisations of gravitational waves.
We consider the equations for scalar modes in considerable detail.
We present the integral solution
of the Boltzmann multipole equations in a $K=0$ almost-FRW universe, and
discuss the relation between line of sight methods (which employ lightlike
integrations along the lightcone) and the Boltzmann multipole
approach (where a timelike integration is performed). We derive
analytic solutions for scalar modes in the early radiation
dominated universe, that are used as initial conditions for the numerical
solution of the scalar equations, the results of which we describe in
Section~\ref{sec_num}. In Section~\ref{sec_tens} we give a brief
discussion of the tensor equations in the covariant approach. The
covariant angular decomposition naturally gives rise to a set of variables
that describe the temperature
anisotropy in a more direct manner than in the conventional metric-based
approaches. This is particularly apparent for tensor perturbations,
where the CMB power spectrum at a given multipole $l$
is determined by the $l-2$, $l$ and $(l+2)$-th moments of the conventional
decomposition of the photon distribution function, which obscures the
physical interpretation of these moments.
Finally, we end with our conclusions in Section~\ref{sec_conc}.
Ultimately, our results confirm those of other
groups (for example, Ma \&\ Bertschinger (1995)) who have performed similar
calculations by
working carefully in specific gauges, but are obtained here with a
unified methodology that is more physically transparent and less prone to
lead to confusion over subtle gauge effects.

We employ standard general relativity and use a $(+---)$ metric signature.
Our conventions for the Riemann and Ricci tensors are fixed by
%
% Preserve subscript/superscript ordering
%
$[\nabla_{a},\nabla_{b}] u^{c} = -{\clr_{abd}}^{c} u^{d}$, and
$\clr_{ab} \equiv {\clr_{acb}}^{c}$. Round brackets around indices denote
symmetrisation on the indices enclosed, and square brackets denote
antisymmetrisation.
We use units with $c=G=1$ throughout, and a unit of distance of $\mpc$ for
numerical work.

\section{The Covariant Approach to Cosmology}
\label{sec_cov}

In this section, we summarise the covariant approach to
cosmology (Ehlers 1993; Ellis 1971; Hawking 1966) and the gauge-invariant
perturbation theory of Ellis \&\ Bruni (1989) which is derived from it.
We begin by choosing a velocity field $u^{a}$ which satisfies the
following criterion: \emph{the velocity must be physically defined such
that it reduces to the four velocity of the fundamental observers in the
Friedmann-Robertson-Walker (FRW) limit}.
This restriction on $u^{a}$ is essential to ensure gauge-invariance of the
Ellis \&\ Bruni perturbation theory. Note that, in a general perturbed
spacetime, there is no unique choice for $u^{a}$. Acceptable choices for
$u^{a}$
include the four velocity of a given matter component, and the timelike
eigenvector of the stress-energy tensor. In the covariant approach to
perturbations in cosmology, covariant variables are introduced that
describe the inhomogeneity and anisotropy of the universe. These
variables employ the velocity field $u^{a}$ in their definition, and so,
in a given spacetime, their values depend on how we choose $u^{a}$
(the exact transformation laws are given in  Maartens et al.\ (1998)).
For a given choice of $u^{a}$, the covariant variables, defined below,
describe the results of observations made by observers comoving with the
velocity $u^{a}$, and their frame-dependence reflects the fact that the
observations depend on the velocity of the observer. It might be thought
that the freedom in the choice of velocity would introduce similar
ambiguities as the choice of map $\Phi$ does in conventional approaches.
However, this is not the case because of the restriction on $u^{a}$ that
we emphasised above. It is certainly true that, with a suitable choice
of $u^{a}$, we can eliminate some aspect of the inhomogeneity and
isotropy observed. For example, we can always choose $u^{a}$ so that the
CMB dipole vanishes. However, in a given spacetime, the covariant variables
cannot be forced to take arbitrary values through some particular choice of
$u^{a}$. In particular, if, for some timelike velocity field (not necessarily
restricted to satisfy the criterion emphasised above), \emph{all} of the
gauge-invariant variables defined below vanish identically, then
the universe is necessarily FRW. This gives a covariant condition which
characterises the FRW limit, but note that, if $u^{a}$ is unrestricted,
we could have the situation where the universe is FRW, but we are not
viewing it from the perspective of the fundamental observers and so
some of the variables would not vanish. (This is similar to
the presence of gauge mode solutions in the metric-based approach.)
However, if we ensure that $u^{a}$ is defined physically, so that in the
FRW limit it necessarily reduces to the velocity of the fundamental observers,
this situation cannot arise, and the variables used to characterise the
anisotropy and inhomogeneity are genuinely gauge-invariant.

We refer to the choice of velocity as a frame choice.
In this paper, we defer making a frame-choice until we have derived all the
relevant equations, so that we have available a set of equations valid
for any choice of $u^{a}$. However, to actually solve the equations,
we must make a definite choice for $u^{a}$ (the system of equations is
under-determined until such a choice is made). Here, it will be convenient
to choose $u^{a}$ to coincide with the velocity of the CDM component,
since $u^{a}$ is then geodesic.

The velocity $u^{a}$ defines
a projection tensor $h_{ab}$ which projects into the space perpendicular to
$u^{a}$ (the instantaneous rest-space of observers moving with velocity
$u^{a}$):
\begin{equation}
h_{ab} \equiv g_{ab} - u_{a}u_{b},
\end{equation}
where $g_{ab}$ is the spacetime metric. Since $h_{ab}$ is a projection
tensor it satisfies
\begin{equation}
h_{ab} = h_{(ab)}, \qquad h_{a}^{c}h_{cb} = h_{ab}, \qquad h^{a}_{a} = 3,
\qquad u^{a} h_{ab} = 0.
\end{equation}

We employ the projection tensor to define a spatial covariant derivative
$\sgrad^{a}$ which acting on a tensor ${T^{b \ldots c}}_{d\ldots e}$
returns a tensor which is orthogonal to $u^{a}$ on every index:
%
% Preserver superscript/subscript order
\begin{equation}
\sgrad^{a} {T^{b \ldots c}}_{d\ldots e} \equiv
h_{p}^{a} h_{r}^{b} \ldots h_{s}^{c} h_{d}^{t}\ldots h_{e}^{u} \nabla^{p}
{T^{r\ldots s}}_{t\ldots u},
\end{equation}
where $\nabla^{a}$ denotes the usual covariant derivative. If the velocity
field $u^{a}$ has vanishing vorticity (see later) $\sgrad^{a}$ reduces to the
covariant derivative in the hypersurfaces orthogonal to $u^{a}$.
 
The covariant derivative of the velocity decomposes as
\begin{equation}
\nabla_{a}u_{b} = \varpi_{ab} + \sigma_{ab} + \tfrac{1}{3}
\theta h_{ab} + u_{a} w_{b},
\end{equation}
where $w_{a} \equiv u^{b}\nabla_{b} u_{a}$ is the acceleration, which satisfies
$u^{a}w_{a}=0$, the scalar $\theta \equiv \nabla^{a} u_{a}=3H$ is the
volume expansion rate ($H$ is the local Hubble parameter),
$\varpi_{ab} \equiv \nabla_{[a}u_{b]} + w_{[a}u_{b]}$ is the vorticity
tensor, which satisfies $\varpi_{ab}=\varpi_{[ab]}$ and $u^{a} \varpi_{ab} =
0$, and $\sigma_{ab} \equiv \sgrad_{(a}u_{b)} - \theta h_{ab} /3$ is the shear
tensor which satisfies $\sigma_{ab}=\sigma_{(ab)}$, $\sigma_{a}^{a} = 0$ and
$u^{a} \sigma_{ab} = 0$. The non-trivial integrability condition
\begin{equation}
\sgrad_{[a}\sgrad_{b]} \phi = -\varpi_{ab} \dot{\phi},
\label{eq_intcon}
\end{equation}
for any scalar field $\phi$, where an overdot denotes the action of the
operator $u^{a}\nabla_{a}$, follows from the Ricci identity.
Note in particular that in an evolving universe ($\dot{\phi}\neq 0$), spatial
gradients are necessarily non-vanishing in the presence of vorticity. This
behaviour, which is a consequence of there being no global hypersurfaces
which are everywhere orthogonal to $u^{a}$ if the vorticity does not
vanish, is central to the discussion of vector perturbations.
For vanishing vorticity, the 3-Ricci scalar (or intrinsic-curvature
scalar) $^{(3)}\clr$ in the hypersurfaces
orthogonal to $u^{a}$ evaluates to
\begin{equation}
^{(3)}\clr = 2\kappa \rho - {\tfrac{2}{3}} \theta^{2} +
\sigma_{ab} \sigma^{ab},
\end{equation}
where $\rho$ is the total energy density in the $u^{a}$ frame.

In an exact FRW universe the vorticity, shear and  acceleration vanish
identically. In an almost-FRW universe, these variables, when suitably
normalised to make them dimensionless, are regarded
as first-order in a smallness parameter $\epsilon$ (Maartens et al. 1995).
We use the convenient notation, $\ord(n)$ to denote that a variable
is $\ord(\epsilon^{n})$. We assume
that products of first-order variables can be neglected from any expression
in the linearised calculation considered here.

Other first-order variables
may be obtained by taking the spatial gradient of scalar quantities. Such
quantities are gauge-invariant by construction since they vanish identically
in an exact FRW universe. We shall make use of the comoving fractional
spatial gradient of the density $\rho^{(i)}$ of a species $i$,
\begin{equation}
\clx_{a}^{(i)} \equiv \frac{S}{\rho^{(i)}} \sgrad_{a} \rho^{(i)},
\end{equation}
and the comoving spatial gradient of the expansion
\begin{equation}
\clz_{a} \equiv S \sgrad_{a} \theta.
\end{equation}
The scalar $S$ is a local scale factor satisfying
\begin{equation}
\dot{S} \equiv u^{a} \nabla_{a} S = H S, \qquad \sgrad^{a} S =\ord(1),
\end{equation}
which removes the effects of the expansion from the spatial gradients
defined above. The vector $\clx_{a}^{(i)}$ is a manifestly covariant and
gauge-invariant characterisation of the density inhomogeneity.
 
The matter stress-energy tensor $\clt_{ab}$ decomposes with respect to
$u^{a}$ as
\begin{equation}
\clt_{ab} \equiv \rho u_{a} u_{b} + 2 u_{(a}q_{b)} - p h_{ab} + \pi_{ab},
\end{equation}
where $\rho\equiv \clt_{ab} u^{a} u^{b}$ is the density of matter (measured
by a comoving observer), $q_{a} \equiv h_{a}^{b} \clt_{bc} u^{c}$ is the
energy (or heat) flux and is orthogonal to $u^{a}$, $p\equiv - h_{ab} \clt^{ab}
/3$ is the isotropic pressure, and the symmetric traceless tensor
$\pi_{ab} \equiv h_{a}^{c} h_{b}^{d} \clt_{cd} + p h_{ab}$ is the anisotropic
stress, which is also orthogonal to $u^{a}$. In an exact FRW universe,
isotropy restricts $\clt_{ab}$ to perfect-fluid form, so that in an almost-FRW
universe the heat flux and isotropic stress may be treated as first-order
variables.
The final first-order gauge-invariant variables we require derive from the Weyl
tensor $\clw_{abcd}$, which vanishes in an exact FRW universe due to isotropy.
The electric and magnetic parts of the Weyl
tensor, denoted by $\cle_{ab}$ and $\clb_{ab}$ respectively, are symmetric
traceless tensors, orthogonal to $u^{a}$, which we define by
\begin{eqnarray}
\cle_{ab} &\equiv&  u^{c}u^{d} \clw_{acbd} \\
\clb_{ab} &\equiv& - \tfrac{1}{2} u^{c} u^{d}{\eta_{ac}}^{ef} \clw_{efbd},
\end{eqnarray}
where $\eta_{abcd}$ is the covariant permutation tensor with $\eta_{0123}=
- \sqrt{-g}$.

\subsection{Linearised Perturbation Equations for the Total Matter Variables}

Exact equations describing the propagation of the total matter variables
(such as the total density $\rho$), the kinematic variables, and the electric
and magnetic parts of the Weyl tensor, and the constraints between them,
follow from the Ricci identity and the Bianchi identity.
The Riemann tensor is expressed in terms of $\cle_{ab}$, $\clb_{ab}$ and the
Ricci tensor, $\clr_{ab}$, and the Einstein equation is used to substitute
for the Ricci tensor in terms of the matter stress-energy tensor. On
linearising the equations that result from this procedure (Bruni, Dunsby, \&\
Ellis 1992),
one obtains five constraint equations:
\begin{eqnarray}
\clb_{ab} + \left(\sgrad^{c} \varpi_{d(a} + \sgrad^{c}\sigma_{d(a}\right)
{\eta_{b)ce}}^{d} u^{e} &=& 0
\label{eq_cons1} \\
\sgrad^{b} \clb_{ab} - \half \kappa \left[ (\rho+p){\eta_{ab}}^{cd} u^{b}
\varpi_{cd} + \eta_{abcd} u^{b} \sgrad^{c} q^{d} \right] &=& 0
\label{eq_cons2} \\
\sgrad^{b} \cle_{ab} - {\tfrac{1}{6}}\kappa \left(2 \sgrad_{a} \rho +
2 \theta q_{a} + 3 \sgrad^{b} \pi_{ab} \right) &=& 0
\label{eq_cons3} \\
\sgrad^{b} \varpi_{ab} + \sgrad^{b} \sigma_{ab} - {\tfrac{2}{3}} \sgrad_{a}
\theta - \kappa q_{a} &=& 0
\label{eq_cons4} \\
\sgrad^{c} \left( \eta_{abcd} u^{d} \varpi^{ab}\right) &=& 0,
\label{eq_cons5}
\end{eqnarray}
and seven propagation equations:
\begin{eqnarray}
\dot{\cle}_{ab} + \theta \cle_{ab} + \sgrad^{c} \clb_{d(a}{\eta_{b)ce}}^{d}
u^{e} + {\tfrac{1}{6}} \kappa \bigl[ 3(\rho+p)\sigma_{ab} && \nonumber\\
+ 3\left(\sgrad_{(a} q_{b)} - {\tfrac{1}{3}}h_{ab} \sgrad^{c} q_{c}\right)
- 3 \dot{\pi}_{ab} - \theta \pi_{ab} \bigr] &=& 0
\label{eq_prop1}\\
\dot{\clb}_{ab} + \theta \clb_{ab} - \left(\sgrad^{c} \cle_{d(a} + \half
\kappa \sgrad^{c} \pi_{d(a} \right) {\eta_{b)ce}}^{d} u^{e} &=& 0
\label{eq_prop2} \\
\dot{\sigma}_{ab} + {\tfrac{2}{3}}\theta \sigma_{ab} - \left(\sgrad_{(a}
w_{b)} - {\tfrac{1}{3}} h_{ab}\sgrad^{c} w_{c} \right) + \cle_{ab} +
\half\kappa \pi_{ab} &=& 0
\label{eq_prop3} \\
\dot{\varpi}_{ab} - \sgrad_{[a}w_{b]} + {\tfrac{2}{3}} \theta \varpi_{ab} &=&0
\label{eq_prop4} \\
\dot{q}_{a} + \tfrac{4}{3} \theta q_{a} + (\rho + p) w_{a} + \sgrad^{b}
\pi_{ab} - \sgrad_{a} p &=& 0
\label{eq_prop5} \\
\dot{\theta} + {\tfrac{1}{3}} \theta^{2}  - \sgrad^{a} w_{a} + \half \kappa
(\rho + 3 p) &=& 0
\label{eq_prop6} \\
\dot{\rho} + \theta (\rho + p) + \sgrad^{a} q_{a} &=& 0,
\label{eq_prop7}
\end{eqnarray}
where $\dot{T}_{ab \ldots c} \equiv u^{d} \nabla_{d} T_{ab \ldots c}$.
The constraint equations do not involve time derivatives, and so they serve
to constrain initial data for the problem. The propagation equations
are consistent with the constraint equations in the sense that the constraints
are preserved in time by the propagation equations if they are satisfied
initially. The consistency of the exact equations follows from their derivation
from the exact field equations, and is preserved by the linearisation
procedure.
Including a cosmological constant $\Lambda$ in the above equations is
straightforward;
one adds a contribution
$\Lambda/\kappa$ to the total density $\rho$, and subtracts the same term from
the total pressure $p$.

Many of the equations given above have simple Newtonian
analogues~\cite{ellis71}, and thus are simple to interpret physically.
The analogues arise because many of the covariantly defined variables
have counterparts in (self-gravitating) Newtonian hydrodynamics. An important
exception is that there is no Newtonian analogue of the magnetic part of the
Weyl tensor (the electric part is analogous to the tidal part of the
Newtonian gravitational potential), and no equation analogous to the
$\dot{\cle}_{ab}$ propagation equation~\cite{ellis97}.
These exceptions arise because
of the instantaneous interaction in Newtonian gravity, which excludes the
possibility of gravitational wave solutions to the Newtonian equations.
However, there is a close analogy between the $\dot{\cle}_{ab}$
and $\dot{\clb}_{ab}$ propagation equations and the
constraint equations~\eqref{eq_cons2} and~\eqref{eq_cons3}, and
Maxwell's equations split with respect to an arbitrary timelike
velocity field (see, for example, Maartens \&\ Bassett (1998)).

There is some redundancy in the full set of linear equations. (See
Maartens (1997) for a discussion of the redundancy in the exact non-linear
equations for an irrotational dust universe.)
For example, equation~\eqref{eq_cons1}, which determines $\clb_{ab}$
in terms of the vorticity and the shear, along with
equation~\eqref{eq_cons4} and the integrability condition
given as equation~\eqref{eq_intcon} imply equation~\eqref{eq_cons2}.
Similarly, equation~\eqref{eq_prop2}
follows from equation~\eqref{eq_cons1} and the propagation equations for the
shear (eq.\ $[\ref{eq_prop3}]$) and the vorticity (eq.\ $[\ref{eq_prop4}]$).
It follows that
$\clb_{ab}$ may be eliminated from the equations in favour
of the vorticity and the shear by making use of equation~\eqref{eq_cons1}. This
elimination is useful when discussing the propagation of vector and tensor
modes (see Section~\ref{sec_tens}).

The usual Friedmann equations describing homogeneous and isotropic
cosmological models are readily obtained from the full set of covariant
equations, since
in an exact FRW universe the only non-trivial propagation equations are the
Raychaudhuri equation (eq.\ [\ref{eq_prop6}]) and the energy conservation
equation (eq.\ [\ref{eq_prop7}]), which reduce to the Friedmann equation
\begin{equation}
\dot{H} + H^{2} = - {\tfrac{1}{6}} \kappa (\rho + 3 p),
\end{equation}
and the usual equation for the density evolution
\begin{equation}
\dot{\rho} = - 3 H (\rho + p).
\end{equation}
The second Friedmann equation is obtained as a first integral of these
two equations:
\begin{equation}
H^{2} + {\tfrac{K}{S^{2}}} = {\tfrac{1}{3}} \kappa \rho,
\end{equation}
where $6 K/S^{2}$ is the intrinsic curvature scalar of the surfaces of
constant cosmic time.

The fractional comoving spatial gradient of the density, $\clx_{a}$, and
the comoving spatial gradient of the expansion rate, $\clz_{a}$, are the key
variables in the covariant discussion of the growth of inhomogeneity in
the universe (Ellis \&\ Bruni 1989; Ellis et al.\ 1989). It is useful to
have available the
propagation equations for these variables. For $\clx_{a}$, we take
the spatial gradient of the density evolution equation (eq.\ [\ref{eq_prop7}])
and commute the space and time derivatives, to obtain
\begin{equation}
\rho \dot{\clx}_{a} + (\rho + p)\left(\clz_{a} - S\theta w_{a}\right)
+ S \sgrad_{a} \sgrad^{b} q_{b} + S \theta \sgrad_{a} p - \theta p \clx_{a}=0.
\end{equation}
For $\clz_{a}$, we take the spatial gradient of the Raychaudhuri
equation (eq.\ [\ref{eq_prop6}]) which gives
\begin{eqnarray}
\lefteqn{\dot{\clz}_{a} + {\tfrac{2}{3}} \theta \clz_{a} - S \left[
{\tfrac{1}{3}} \theta^{2} + {\tfrac{1}{2}}\kappa (\rho + 3 p)\right]w_{a}
+ {\tfrac{1}{2}} \kappa
S \left( \sgrad_{a} \rho + 3 \sgrad_{a} p \right)}\hspace{+8cm} \nonumber \\
&&\mbox{} - S \sgrad_{a} \sgrad^{b} w_{b} = 0,
\end{eqnarray}
For an ideal fluid ($q_{a} = \pi_{ab} = 0$ when
we choose $u^{a}$ to be the fluid velocity), with a barotropic equation of
state $p=p(\rho)$, the propagation equations for $\clx_{a}$ and
$\clz_{a}$ combine with the momentum conservation equation (eq.\
[\ref{eq_prop5}]) and the integrability condition, given as
equation~\eqref{eq_intcon}, to give an inhomogeneous
second-order equation for $\clx_{a}$~\cite{ellis90}. For a simple equation
of state $p=(\gamma-1)\rho$, where $\gamma$ is a constant, the second-order
equation is
\begin{eqnarray}
\lefteqn{\ddot{\clx}_{a} + \left({\tfrac{5}{3}}-\gamma\right)
\theta \dot{\clx}_{a}
+ {\tfrac{1}{2}} (\gamma-2)(3\gamma -2) \left( {\tfrac{1}{3}}\theta^{2} +
{\tfrac{3K}{S^{2}}} \right) \clx_{a} } \nonumber \\
&&\mbox{}+ (\gamma-1) \left(\sgrad^{2} \clx_{a}
+ {\tfrac{2K}{S^{2}}} \clx_{a} \right) + 2\gamma (\gamma-1) S \theta
\sgrad^{b} \varpi_{ab} = 0.
\label{eq_clxdtdt}
\end{eqnarray}
From this equation, it is straightforward to recover the usual results
for the growth of inhomogeneities in an almost-FRW universe~\cite{ellis89a}.
The inhomogeneous term describes the coupling between the vorticity and the
spatial gradient of the density, which arises from the lack of global
hypersurfaces orthogonal to $u^{a}$ in the presence of non-vanishing vorticity.
In reality, the universe cannot be described by a barotropic perfect fluid.
A more careful analysis of the individual matter components is required,
which we present in the next section.

\section{Equations for Individual Matter Components}
\label{sec_eqs}

In this paper we concentrate on CDM models, so the matter components that we
must consider are the photons and neutrinos, which are the only
relativistic species, and the tightly-coupled baryon/electron system and the
CDM, which are both non-relativistic over the epoch of interest.
We consider the description of each of these components separately in this
section.

\subsection{Photons}

In relativistic kinetic theory (see, for example, Misner, Thorne, \&\
Wheeler (1973)), the
photons are described by a scalar-valued distribution function
$\fgam(x,p)$.
An observer sees $\fgam(x,p) d^{3}x d^{3}p$ photons at the
spacetime point $x$ in a proper volume $d^{3}x$, with covariant momentum
$p^{a}$ in a proper volume $d^{3}p$ of momentum space. The photon momentum
$p^{a}$ decomposes with respect to the velocity $u^{a}$ as
\begin{equation}
p^{a} = E (u^{a} + e^{a}),
\label{eq_psplit}
\end{equation}
where $E = p^{a} u_{a}$ is the energy of the photon, as measured by an observer
moving with velocity $u^{a}$, and $e_{a}$ is a unit spacelike vector which is
orthogonal to $u^{a}$:
\begin{equation}
e^{a}e_{a} = -1, \qquad e^{a} u_{a} = 0,
\end{equation}
which describes the propagation direction of the photon in the instantaneous
rest space of the observer. With this decomposition of the momentum, we may
write the photon distribution function in the form $\fgam(E,e)$
when convenient, where
the dependence on spacetime position $x$ has been left implicit.
The stress-energy tensor $\clt^{(\gamma)}_{ab}$
for the photons may then be written as
\begin{equation}
\clt^{(\gamma)}_{ab} = \int dE d\Omega \, E \fgam(E,e) p_{a} p_{b},
\label{eq_radset}
\end{equation}
where the measure $d\Omega$ denotes an integral over solid angles.
The photon energy density $\rho^{(\gamma)}$, the heat flux $q^{(\gamma)}_{a}$,
and the anisotropic stress $\pi_{ab}^{(\gamma)}$ are given by integrals of the
three lowest moments of the photon distribution function:
\begin{eqnarray}
\rho^{(\gamma)} &=& \int dE d\Omega \, E^{3} \fgam(E,e) \label{eq_mom1}\\
q_{a}^{(\gamma)} &=& \int dE d\Omega \, E^{3} \fgam(E,e) e_{a} \\
\pi_{ab}^{(\gamma)} &=& \int dE d\Omega \, E^{3} \fgam(E,e) e_{a} e_{b}
+ {\tfrac{1}{3}} \rho^{(\gamma)} h_{ab}.\label{eq_mom3}
\end{eqnarray}

In the absence of scattering, the photon distribution is conserved in
phase space. Denoting the photon position by $x^{a}(\lambda)$
and the momentum by $p^{a}(\lambda)$, the path in phase space is described by
the equations
\begin{eqnarray}
\frac{dx^{a}}{d\lambda} &=& p^{a} \\
p^{a} \nabla_{a} p^{b} &=& 0,
\end{eqnarray}
where $\lambda$ is an affine parameter along the null geodesic
$x^{a}(\lambda)$.
Denoting the Liouville operator by $\cll$, we have
\begin{equation}
\cll \fgam(x,p) = {\frac{d}{d\lambda}} \fgam(x^{a}(\lambda),p^{a}(\lambda))
= 0,
\end{equation}
in the absence of collisions. Over the epoch of interest here, the photons
are not collisionless, but instead are interacting with a thermal distribution
of electrons and baryons. The dominant contribution to the scattering comes
from Compton scattering off free electrons, which have number density
$\nelec$ in the baryon/electron rest frame. Since the average energy of a CMB
photon is small compared to the electron mass well after electron-positron
annihilation, we may approximate the Compton scattering by Thomson scattering.
Furthermore, since the kinetic temperature of the electrons (which equals the
radiation temperature prior to recombination) is small compared to the
electron mass, the electrons are non-relativistic and we may ignore the
effects of thermal motion of the electrons (in the average rest frame of the
baryon/electron system) on the scattering. Our final assumption is to
ignore polarisation of the radiation. Thomson scattering of an unpolarised
but anisotropic distribution of radiation leads to the generation of
polarisation, which then affects the temperature anisotropy because of the
polarisation dependence of the Thomson cross section $\sigma_{T}$. In this
manner, polarisation of the CMB is generated through recombination and its
neglect leads to errors of a few percent~\cite{hu95} in the predicted
temperature anisotropy. We hope to develop a covariant version of the
radiative transfer equations including polarisation in the near future,
which should simplify their physical interpretation.

In the presence of scattering, the photon distribution function evolves
according to the collisional Boltzmann equation,
\begin{equation}
\cll \fgam(x,p) = \clc,
\end{equation}
where the collision operator for Thomson scattering is
\begin{equation}
\clc = \nelec \sigma_{T} p^{a} u_{a}^{(b)} \left[\fgam_{+}(x,p)-
\fgam(x,p)\right],
\label{eq_clc}
\end{equation}
where $u_{a}^{(b)}$ is the covariant velocity of the baryon/electron system
and $\fgam_{+}(x,p)$ describes scattering into the phase space element under
consideration:
\begin{equation}
\fgam_{+}(x,p) = \frac{3}{16\pi} \int \fgam(x,p') \left[1 + \left(g^{ab}
e^{(b)}_{a} e^{\prime (b)}_{b}\right)^{2} \right] \, d\Omega_{e^{\prime (b)}},
\end{equation}
where $e^{(b)}_{a}$ is the photon direction relative to $u^{(b)}_{a}$,
\begin{equation}
p_{a} = E^{(b)} \left(u^{(b)}_{a} + e^{(b)}_{a}\right), \qquad E^{(b)}
= p^{a} u^{(b)}_{a}, \label{eq_psplitb}
\end{equation}
and $e^{\prime(b)}_{a}$ is the initial direction
(relative to $u^{(b)}_{a}$) of the photon whose initial momentum is
$p'_{a}$ and final momentum is $p_{a}$.
We write the baryon velocity in the form
\begin{equation}
u^{(b)}_{a} = \gamma^{(b)} \left(u_{a} + v^{(b)}_{a}\right),
\end{equation}
where $v_{a}^{(b)}$ is the relative velocity of the baryons,
which satisfies $u^{a}v_{a}^{(b)}=0$, and $\gamma^{(b)}\equiv
(1+g^{ab} v_{a}^{(b)}v_{b}^{(b)})^{-1/2}$. Note that to first-order
we have $u_{a}^{(b)}=u_{a} + v_{a}^{(b)}$ since the relative velocities
of the individual matter components are first-order in an almost-FRW universe.
Multiplying the Boltzmann equation by $E^{2}$ and integrating over energies,
we find
\begin{eqnarray}
\int dE\, E^{2}\cll f^{(\gamma)}(E,e) &=& \nelec\sigma_{T}\left[\gamma^{(b)}
\left(1+e^{a}v_{a}^{(b)}\right)\right]^{-3} \int dE^{(b)}\, {E^{(b)}}^{3}
f_{+}^{(\gamma)}(x,p)\nonumber \\
&&\mbox{} - \nelec \sigma_{T}\gamma^{(b)} \left(1+e^{a}v_{a}^{(b)}\right)
\int dE\, E^{3} f^{(\gamma)}(E,e),
\end{eqnarray}
where we have used
\begin{equation}
E^{(b)} = \gamma^{(b)} E \left(1+e^{a}v_{a}^{(b)}\right),
\end{equation}
to replace the integral over $E$ by an integral over $E^{(b)}$ in the
first-term on the right. This term can be rewritten as an integral over
$E^{\prime(b)} \equiv p^{\prime a}u_{a}^{(b)}$ using the fact that there is
no energy transfer in Thomson scattering in the rest frame of the
scattering electron, so that
\begin{equation}
\int dE^{(b)}\, {E^{(b)}}^{3} f_{+}^{(\gamma)}(x,p) = \frac{3}{16\pi}
\int dE^{\prime (b)} d\Omega_{e^{\prime (b)}} {E^{\prime(b)}}^{3}
\left[1+ \left(g^{ab} e_{a}^{(b)} e_{b}^{\prime(b)} \right)^{2}\right]
f^{(\gamma)}(x,p).
\end{equation}
Using the definition of the radiation stress-energy tensor
(eq.~\eqref{eq_radset}) in the right-hand side, we have
\begin{equation}
\int dE^{(b)}\, {E^{(b)}}^{3} f_{+}^{(\gamma)}(x,p) =
\frac{3}{16\pi} g^{ab}g^{cd}\clt_{bd}^{(\gamma)}
\left( u_{a}^{(b)}u_{c}^{(b)} + e_{a}^{(b)} e_{c}^{(b)}\right),
\end{equation}
where $e^{(b)}_{a}$ can be expressed as
\begin{equation}
e_{a}^{(b)} = \left[\gamma^{(b)}\left(1+e^{c}v_{c}^{(b)}\right)\right]^{-1}
(u_{a} + e_{a}) - \gamma^{(b)} \left(u_{a} + v_{a}^{(b)}\right).
\end{equation}
It follows that the energy-integrated Boltzmann equation reduces to
\begin{eqnarray}
\int dE \, E^{2} \cll \fgam(E,e) &=& {\frac{3}{16\pi}} \nelec\sigma_{T}
\left[\gamma^{(b)}\left(1+e^{f}v_{f}^{(b)}\right)\right]^{-3}
g^{ab}g^{cd}\clt_{bd}^{(\gamma)}
\left( u_{a}^{(b)}u_{c}^{(b)} + e_{a}^{(b)} e_{c}^{(b)}\right)
\nonumber \\
&&\mbox{} - \nelec \sigma_{T} \gamma^{(b)} \left(1+e^{a}v_{a}^{(b)}\right)
\int dE\, E^{3} f^{(\gamma)}(E,e).
\label{eq_bolex}
\end{eqnarray}
This equation is exact under the assumption of Thomson scattering and the
neglect of polarisation. Here, we shall only require the linearised version
of equation~\eqref{eq_bolex}; for a covariant discussion of the second-order
effects in this equation, see Maartens et al.\ (1998).
On linearising equation~\eqref{eq_bolex} around an almost-FRW universe,
we find
\begin{eqnarray}
\lefteqn{\int dE \, E^{2} \cll \fgam(E,e) = {\frac{3}{16\pi}} \nelec\sigma_{T}
\left[{\frac{4}{3}}\left(1-4e^{a}v_{a}^{(b)}\right)\rho^{(\gamma)}
+ \pi_{ab}^{(\gamma)} e^{a}e^{b} \right]}\hspace{8cm}\nonumber
\\
&&\mbox{} - \nelec \sigma_{T} \int dE\, E^{3} \fgam(E,e).
\label{eq_boltz}
\end{eqnarray}
This covariant form of the Boltzmann equation was used in Challinor
\&\ Lasenby (1998) to discuss CMB anisotropies from scalar perturbations on
angular scales
above the damping scale. Note that the equation is fully covariant with
all variables observable in the real universe, is valid for arbitrary type
of perturbation (scalar, vector and tensor), employs no harmonic
decomposition and is valid for any background FRW model.

The numerical solution of the Boltzmann equation (eq.~\eqref{eq_boltz})
is greatly
facilitated by decomposing the equation into covariantly-defined angular
moments. The majority of recent calculations (for example, Seljak \&\
Zaldarriaga (1996))
perform an angular decomposition of the Boltzmann equation after specifying
the perturbation type and performing the appropriate harmonic expansions.
The procedure is straightforward for scalar perturbations in a $K=0$
universe, where the Fourier mode of the perturbation in the distribution
function may be assumed to be axisymmetric about the wavevector $\bk$
(this assumption is consistent with the evolution implied by the Boltzmann
equation), allowing an angular expansion in Legendre polynomials alone.
However, for tensor perturbations the situation is not so straightforward
(see, for example, Kosowsky (1996)), since the Boltzmann equation does
not then support axisymmetric modes.
Instead, the necessary azimuthal dependence
of the Fourier components of the perturbation in the distribution function,
which is different for the two polarisations of the tensor modes, must be put
in by hand, prior to a Legendre expansion in the polar angle. This procedure
may be eliminated by performing a covariant angular expansion of
$\fgam(x,p)$ prior to specifying the perturbation type or background FRW model.
The covariant (tensor) moment equations that result may then be solved for
any type of perturbation (and any background curvature $K$) by expanding in
covariant tensors derived from the appropriate harmonic
functions (see Section~\ref{sec_scal} for the case of scalar perturbations
and Section~\ref{sec_tens} for tensor perturbations).
This procedure automatically takes care of the required angular dependencies
of the harmonic components of the distribution function, allowing a streamlined
and unified treatment of all perturbation types in background FRW models
with arbitrary spatial curvature.

The covariant angular expansion of the photon distribution function takes the
form (Ellis, Matravers, \&\ Treciokas 1983, Thorne 1981)
\begin{equation}
\fgam(E,e) = \sum_{l=0}^{\infty} \Fgaml_{a_{1}\ldots a_{l}} e^{a_{1}}
e^{a_{2}} \ldots e^{a_{l}},
\label{eq_exp}
\end{equation}
where the tensors $\Fgaml_{a_{1}\ldots a_{l}}$ have an implicit
dependence on spacetime position $x$ and energy $E$, and are totally symmetric,
traceless and orthogonal to $u^{a}$:
\begin{equation}
\Fgaml_{a_{1}\ldots a_{l}} =
\Fgaml_{(a_{1}\ldots a_{l})}, \quad
g^{a_{1}a_{2}} \Fgaml_{a_{1} a_{2} \ldots a_{l}} = 0, \quad
u^{a_{1}} \Fgaml_{a_{1} \ldots a_{l}}  = 0.
\end{equation}
Employing the expansion given in equation~\eqref{eq_exp},
the action of the Liouville operator on $\fgam(E,e)$ reduces to
\begin{equation}
\cll \fgam(E,e) = \sum_{l=0}^{\infty} \left[\partial_{E}
\Fgaml_{a_{1}\ldots a_{l}} e^{a_{1}} \partial_{\lambda} E + p^{b} \nabla_{b}
\Fgaml_{a_{1}\ldots a_{l}} e^{a_{1}} + l \Fgaml_{a_{1}\ldots a_{l}}
p^{b} \nabla_{b} e^{a_{1}} \right] e^{a_{2}}\ldots e^{a_{l}}.
\label{eq_liou1}
\end{equation}
Using the geodesic equation, we find that
\begin{equation}
h_{ab} p^{c} \nabla_{c} e^{b} = - E\left(\sigma_{bc}e^{b}e^{c}e_{a}
+ \sigma_{ab}e^{b} + w^{b}e_{a}e_{b} + w_{a} - \varpi_{ab} e^{b}\right),
\end{equation}
which is first-order. In an exact FRW universe, isotropy restricts
$\Fgaml_{a_{1}\ldots a_{l}} = 0$ for $l > 0$, so that in an almost-FRW
universe $\Fgaml_{a_{1}\ldots a_{l}} = \ord(1)$ for $l$ not equal to zero.
It follows that the last term in equation~\eqref{eq_liou1} makes only a
second-order contribution and may be dropped in the linear calculation
considered here.

Inserting the expansion given in equation~\eqref{eq_exp} into the Boltzmann
equation (eq.\ [\ref{eq_boltz}]) and performing a covariant angular expansion
of the resulting equation gives a set of moment equations which are
equivalent to the original Boltzmann equation. The linearised calculation
is straightforward, although a little care is needed for the first three
moments since $\Fgamlin{0}$ is a zero-order quantity. (The exact expansion
of the left-hand side of the Boltzmann equation, equation~\eqref{eq_liou1},
is given in Ellis et al.\ (1983) and Thorne (1981).) For $l=0$, $1$ and $2$,
we find
\begin{eqnarray}
\dot{\rho}^{(\gamma)} + {\tfrac{4}{3}} \theta \rho^{(\gamma)} +
\sgrad^{a} q_{a}^{(\gamma)} &=& 0 \label{eq_rhogamdt} \\
\dot{q}_{a}^{(\gamma)} + {\tfrac{4}{3}} \theta q_{a}^{(\gamma)} +
\sgrad^{b} \pi_{ab}^{(\gamma)} + {\tfrac{4}{3}} \rho^{(\gamma)} w_{a}
- {\tfrac{1}{3}} \sgrad_{a} \rho^{(\gamma)} &=& \nelec \sigma_{T}
\left({\tfrac{4}{3}} \rho^{(\gamma)} v_{a}^{(b)} - q_{a}^{(\gamma)} \right)\\
\lefteqn{\dot{\pi}^{(\gamma)}_{ab} + {\tfrac{4}{3}} \theta \pi^{(\gamma)}_{ab}
+ \sgrad^{c} \Jgamlin{3}_{abc} - {\tfrac{2}{5}} \left(
\sgrad_{(a}q_{b)}^{(\gamma)} - {\tfrac{1}{3}} h_{ab} \sgrad^{c}
q_{c}^{(\gamma)} \right) - {\tfrac{8}{15}} \rho^{(\gamma)}\sigma_{ab}}
\hspace{8cm}\nonumber\\
&=& - {\tfrac{9}{10}} \nelec \sigma_{T} \pi^{(\gamma)}_{ab},
\label{eq_mompi}
\end{eqnarray}
and, for $l \geq 3$,
\begin{eqnarray}
\lefteqn{\dot{J}^{(l)}_{a_{1}\ldots a_{l}} + {\tfrac{4}{3}}\theta
\Jgaml_{a_{1}\ldots a_{l}} + \sgrad^{b} \Jgamlin{l+1}_{b a_{1}\ldots a_{l}}
- {\tfrac{l}{(2l+1)}} \Bigl(\sgrad_{(a_{1}} \Jgamlin{l-1}_{a_{2}\ldots a_{l})}}
\nonumber \\
&& \mbox{}- {\tfrac{(l-1)}{(2l-1)}} \sgrad^{b} \Jgamlin{l-1}_{b (a_{1}
\ldots a_{l-2}} h_{a_{l-1}a_{l})} \Bigr) = - \nelec \sigma_{T}
\Jgaml_{a_{1}\ldots a_{l}}.
\label{eq_moml}
\end{eqnarray}
The tensors $\Jgaml_{a_{1}\ldots a_{l}}$, which are traceless, totally
symmetric and orthogonal to $u^{a}$, are derived from the $\Fgaml_{a_{1}\ldots
a_{l}}$ by integrating over energy:
\begin{equation}
\Jgaml_{a_{1}\ldots a_{l}} \equiv {\frac{4\pi (-2)^{l} (l!)^{2}}{(2l+1)(2l)!}}
\int_{0}^{\infty} dE \, E^{3} \Fgaml_{a_{1}\ldots a_{l}}.
\end{equation}
The constant factor is chosen to simplify algebraic factors in the moment
equations. Using equations~(\ref{eq_mom1}--\ref{eq_mom3}), the lowest
three moments relate simply to the energy density, heat flux and anisotropic
stress:
\begin{equation}
\rho^{(\gamma)} = \Jgamlin{0}, \quad q^{(\gamma)}_{a} = \Jgamlin{1}_{a},
\quad \pi^{(\gamma)}_{ab} = \Jgamlin{2}_{ab}.
\end{equation}
It is straightforward to show that the tensor
\begin{equation}
\sgrad_{(a_{1}}\Jgamlin{l-1}_{a_{2}\ldots a_{l})} - {\tfrac{(l-1)}{(2l-1)}}
\sgrad^{b} \Jgamlin{l-1}_{b (a_{1} \ldots a_{l-2}} h_{a_{l-1}a_{l})},
\end{equation}
which appears in equation~\eqref{eq_moml} is traceless, symmetric, and
orthogonal to $u^{a}$, as required.

It will be observed that for $l\geq 3$, the moment equations link the
$l-1$, $l$ and $l+1$ angular moments of the (integrated) distribution
function, while the $l=2$ equation also involves the density $\rho^{(\gamma)}$
which is the $l=0$ moment. The exact moment equations that arise from
expanding the Liouville equation in covariant harmonics also couple the
$l+2$ and $l-2$ moments to $\Jgaml_{a_{1}\ldots a_{l}}$ (Ellis et al.\ 1983),
but these terms are second-order for $l\geq 3$ and so do not appear in the
linearised equations presented here. In the exact expansion of the Liouville
equation, the coefficient of the $l+2$ angular moment in the exact
propagation equation for $\Jgaml_{a_{1}\ldots a_{l}}$ is the shear
$\sigma_{ab}$, which leads to the result that the angular expansion of the
distribution function for non-interacting radiation can only truncate
($\Jgaml_{a_{1}\ldots a_{l}}=0$ for all $l$ greater than some $L$) if the
shear vanishes~\cite{ellis-er96}. This exact result, which is lost in
linearised theory which permits truncated distribution functions with
non-vanishing shear, is an example of a linearisation instability
(see Ellis \&\ Dunsby (1997b) for more examples).
However, this is not problematic
for the linearised calculation of CMB anisotropies since it is never claimed
that the higher-order moments of the photon distribution vanish exactly.
Instead, the series is truncated (with suitable care to avoid reflection of
power back down the series) for numerical convenience. The truncation is
performed with $L$ large enough so that there is no significant effect on
the $\Jgaml_{a_{1}\ldots a_{l}}$ for the range of $l$ of interest.

Finally, by taking the spatial gradient of equation~\eqref{eq_rhogamdt}, and
commuting the space and time derivatives, we find the propagation equation
for the comoving
fractional spatial gradient of the photon density, $\clx^{(\gamma)}_{a}$,
\begin{equation}
\dot{\clx}^{(\gamma)}_{a} + {\tfrac{4}{3}} \clz_{a} +
{\tfrac{S}{\rho^{(\gamma)}}} \sgrad^{a} \sgrad^{b} q^{(\gamma)}_{b}
- {\tfrac{4}{3}} S \theta w_{a} =0,
\end{equation}
where $\clz_{a}$ is the comoving spatial gradient of the volume expansion.

\subsection{Neutrinos}

We consider only massless neutrinos, and these are non-interacting over the
epoch of interest. It follows that their distribution function
$f^{(\nu)}(x,p)$ satisfies the Liouville equation $\cll f^{(\nu)}(x,p)=0$.
Expanding the neutrino distribution function in covariant angular harmonics,
we arrive at the moment equations for the tensors $\Gnul_{a_{1}\ldots a_{l}}$,
which are defined in the same manner as the
$\Jgaml_{a_{1}\ldots a_{l}}$, but with the photon distribution function
replaced by the neutrino distribution. These moment equations are the same
as the photon equations, but with the scattering terms omitted:
\newpage
\begin{eqnarray}
\dot{\rho}^{(\nu)} + {\tfrac{4}{3}}\theta \rho^{(\nu)} + \sgrad^{a}
q^{(\nu)}_{a} &=& 0 \label{eq_rhonudt} \\
\dot{q}^{(\nu)}_{a} + {\tfrac{4}{3}}\theta q_{a}^{(\nu)} +
\sgrad^{b} \pi^{(\nu)}_{ab} + {\tfrac{4}{3}} \rho^{(\nu)} - {\tfrac{1}{3}}
\sgrad^{a} \rho^{(\nu)} &=& 0 \\
\dot{\pi}^{(\nu)}_{ab} + {\tfrac{4}{3}}\theta \pi^{(\nu)}_{ab} + \sgrad^{c}
\Gnulin{3}_{abc} - {\tfrac{2}{5}} \left( \sgrad_{(a} q^{(\nu)}_{b)} -
{\tfrac{1}{3}} h_{ab} \sgrad^{c} q_{c}^{(\nu)} \right) - {\tfrac{8}{15}}
\rho^{(\nu)} \sigma_{ab}  &=& 0,
\end{eqnarray}
and for $l\geq 3$,
\begin{eqnarray}
\lefteqn{\dot{G}^{(l)}_{a_{1}\ldots a_{l}} + {\tfrac{4}{3}} \theta
\Gnul_{a_{1}\ldots a_{l}} + \sgrad^{b} \Gnulin{l+1}_{b a_{1}\ldots a_{l}} -
{\tfrac{l}{(2l+1)}} \Bigl( \sgrad_{(a_{1}} \Gnulin{l-1}_{a_{2}\ldots a_{l})}}
\hspace{5cm}\nonumber \\
&& \mbox{}- {\tfrac{(l-1)}{(2l-1)}} \sgrad^{b} \Gnulin{l-1}_{b (a_{1}
\ldots a_{l-2}} h_{a_{l-1} a_{l})} \Bigr)= 0.
\end{eqnarray}
The propagation equation for the comoving fractional spatial gradient
of the neutrino density, $\clx^{(\nu)}_{a}$, follows from
equation~\eqref{eq_rhonudt}:
\begin{equation}
\dot{\clx}^{(\nu)}_{a} + {\tfrac{4}{3}} \clz_{a} + {\tfrac{S}{\rho^{(\nu)}}}
\sgrad^{a} \sgrad^{b} q^{(\nu)}_{b} - {\tfrac{4}{3}} S \theta w_{a} = 0.
\end{equation}

\subsection{Baryons}

Over the epoch of interest here, the electrons and baryons are
non-relativistic, and may be approximated by a tightly-coupled ideal fluid
(the coupling arising from Coulomb scattering). The energy density of the
fluid is $\rho^{(b)}$, which includes contributions from both the baryonic
species and the electrons, the fluid pressure is $p^{(b)}$, and
the velocity of the fluid is $u^{(b)}_{a} = u_{a} + v^{(b)}_{a}$ to
first-order, where the $\ord(1)$ relative
velocity $v^{(b)}_{a}$ satisfies $u^{a} v^{(b)}_{a} = \ord(2)$.

The linearised baryon stress-energy tensor evaluates to
\begin{equation}
\clt^{(b)}_{ab} = \rho^{(b)} u_{a} u_{b} - p^{(b)} h_{ab}
+ 2 (\rho^{(b)} + p^{(b)}) u_{(a}v^{(b)}_{b)},
\end{equation}
which shows that there is a heat flux $(\rho^{(b)} + p^{(b)}) v^{(b)}_{a}$
due to the baryon motion relative to the $u^{a}$ frame. The equations of motion
for $\rho^{(b)}$ and $v^{(b)}_{a}$ follow from the conservation of baryon
plus photon stress-energy (the baryons and photons interact through
non-gravitational effects only with themselves):
\begin{equation}
\nabla^{a} \clt^{(b)}_{ab} + \nabla^{a} \clt^{(\gamma)}_{ab} = 0.
\end{equation}
Using the $l=0$ and $l=1$ moment equations for the photon distribution, we find
the propagation equation for the baryon energy density:
\begin{equation}
\dot{\rho}^{(b)} + (\rho^{(b)} + p^{(b)}) \theta + (\rho^{(b)} + p^{(b)})
\sgrad^{a} v_{a}^{(b)} = 0,
\label{eq_rhobdt}
\end{equation}
and a propagation equation for $v^{(b)}_{a}$:
\begin{eqnarray}
\lefteqn{(\rho^{(b)} + p^{(b)}) (\dot{v}_{a}^{(b)} + w_{a}) + {\tfrac{1}{3}}
(\rho^{(b)} + p^{(b)}) \theta v^{(b)}_{a} + \dot{p}^{(b)} v^{(b)}_{a}
- \sgrad_{a} p^{(b)}} \hspace{6cm}\nonumber \\
&& \mbox{}+ \nelec \sigma_{T} \left({\tfrac{4}{3}} \rho^{(\gamma)}
v^{(b)}_{a} - q^{(\gamma)}_{a} \right)=0,
\label{eq_vbdt}
\end{eqnarray}
which must be supplemented by an equation of state linking $p^{(b)}$ and
$\rho^{(b)}$. The final term in equation~\eqref{eq_vbdt} describes the exchange
of momentum between the radiation and the baryon/electron fluid as
a result of Thomson scattering. There is no such term in
equation~\eqref{eq_rhobdt}
since both the radiation drag force and the baryon velocity relative to the
$u^{a}$ frame are first-order, which give only a second-order rate of energy
transfer in the $u^{a}$ frame. Energy transfer due to thermal motion of the
electrons in the baryon rest frame has negligible effect on $\rho^{(b)}$
since the electrons are non-relativistic; $k_{B} T^{(b)} \ll m_{e}$, where
$T^{(b)}$ is the baryon kinetic temperature (assumed equal to the electron
kinetic temperature), and $m_{e}$ is the electron mass.

Taking the spatial gradient of equation~\eqref{eq_rhobdt} gives the
propagation equation
for $\clx_{a}^{(b)}$, the fractional comoving spatial gradient of the baryon
energy density:
\begin{eqnarray}
\lefteqn{\rho^{(b)} \dot{\clx}_{a}^{(b)} + \left(\rho^{(b)} + p^{(b)}\right)
\left(\clz_{a} + S \sgrad_{a} \sgrad^{b} v^{(b)}_{b} - S \theta w_{a}\right)}
\hspace{5cm}\nonumber \\
&&\mbox{} + S \theta \sgrad_{a} p^{(b)} - \theta p^{(b)} \clx_{a}^{(b)}=0.
\end{eqnarray}
We have retained all terms involving the baryon pressure $p^{(b)}$ in the
equations of this section. In practice, over epochs where the baryons are
non-relativistic ($p^{(b)} \ll \rho^{(b)}$), the only pressure term that
need be retained is the term $\sgrad^{a} p^{(b)}$ which appears
in equation~\eqref{eq_vbdt}.
This term appears as a small correction to the total
sound speed in the tightly-coupled baryon/photon plasma, and is potentially
significant during the acoustic oscillations in the plasma.

\subsection{Cold Dark Matter}

We will only consider cold dark matter (CDM) here, which may be described
as a pressureless ideal fluid. Hot dark matter (HDM), which requires a phase
space description, is considered in Ma \&\ Bertschinger (1995),
where the calculations for scalar perturbations are performed in the
synchronous and conformal Newtonian gauges. The CDM has energy density
$\rho^{(c)}$ in its rest frame, which has velocity $u^{(c)}_{a} = u_{a}
+ v^{(c)}_{a}$, with the first-order relative velocity $v^{(c)}_{a}$
satisfying $u^{a} v_{a}^{(c)} = \ord(2)$. The linearised
CDM stress-energy tensor evaluates to
\begin{equation}
\clt^{(c)}_{ab} = \rho^{(c)} u_{a} u_{b} + 2 \rho^{(c)} u_{(a}v_{b)}^{(c)},
\end{equation}
which is conserved since the CDM interacts with other species only through
gravity. The conservation of $\clt^{(c)}_{ab}$ gives the propagation
equations for $\rho^{(c)}$ and $v^{(c)}_{a}$:
\begin{eqnarray}
\dot{\rho}^{(c)} + \theta \rho^{(c)} + \rho^{(c)} \sgrad^{a} v_{a}^{(c)}
&=& 0 \label{eq_rhocdt}\\
\dot{v}_{a}^{(c)} + {\tfrac{1}{3}} \theta v_{a}^{(c)} + w_{a} &=& 0.
\end{eqnarray}
Since the CDM moves on geodesics, the velocity $u^{(c)}_{a}$ provides a
convenient frame choice. With this choice, the acceleration $w_{a}$ vanishes.
We use the CDM frame to define the fundamental velocity $u^{a}$ in
Section~\ref{sec_scal}, where we discuss scalar perturbations in the CDM
model. For the moment, however, we continue to leave the choice of
frame unspecified for generality.
The final equation that we require is the propagation equation for the
fractional comoving spatial gradient of the density, $\clx^{(c)}_{a}$.
This follows from equation~\eqref{eq_rhocdt} on taking the spatial gradient:
\begin{equation}
\dot{\clx}_{a}^{(c)} + \clz_{a} + S \sgrad_{a} \sgrad^{b} v^{(c)}_{b}
- S \theta w_{a} = 0.
\end{equation}

The equations for the matter components that we have described in this section
combine with the covariant equations of Section~\ref{sec_cov} to give
a complete description of the evolution of inhomogeneity and anisotropy
in a fully covariant and gauge-invariant manner. The equations given in
Section~\ref{sec_cov} make use of the total energy density and pressure,
heat flux and anisotropic stress. These quantities are related to the
individual matter components in the CDM model by
\begin{eqnarray}
\rho &=& \rho^{(\gamma)} +\rho^{(\nu)} + \rho^{(b)} + \rho^{(c)} \\
p &=& {\tfrac{1}{3}} \rho^{(\gamma)} +   {\tfrac{1}{3}}\rho^{(\nu)}
+ p^{(b)}\\
q_{a} &=& q_{a}^{(\gamma)} +  q_{a}^{(\nu)} + \left(\rho^{(b)} + p^{(b)}\right)
v_{a}^{(b)} + \rho^{(c)} v_{a}^{(c)} \\
\pi_{ab} &=& \pi^{(\gamma)}_{ab} + \pi^{(\nu)}_{ab}.
\end{eqnarray}

The equations given here are both covariant and gauge-invariant. Employing
gauge-invariant variables ensures that the problem of gauge-mode solutions does
not arise, and that all quantities are independent of the choice of map
between the real universe and a background FRW model. We have only considered
the linearised equations here, but the linearisation procedure is
not fundamental to the covariant and gauge-invariant approach. It
is straightforward to extend the treatment to include non-linear
effects (Maartens et al.\ 1998), which should provide a systematic footing for
the discussion of second-order effects on the CMB.
Indeed, the simplicity with which exact,
non-linear equations can be written down and manipulated is a significant
virtue of the covariant approach.
Unlike in Bardeen's gauge-invariant approach~\cite{bardeen80}, the definition
of the variables employed here does not require that the perturbations be
in the linear regime, and furthermore, the variables do not depend on
the non-local decomposition of the perturbations into scalar, vector and tensor
type and the associated harmonic analysis. The covariant approach describes
scalar, vector and tensor modes in a unified manner, although decomposing
the linear perturbations is useful to aid solution of the linearised equations
late on in the calculation. A further advantage of the covariant and
gauge-invariant approach over that introduced by Bardeen, is that only
covariantly defined variables are employed, which are simple to interpret
physically. In contrast, the Bardeen variables are constructed by taking
linear combinations of (gauge-dependent) metric and matter perturbations
in such a way that the resulting variable is gauge-invariant (for small
gauge-transformations which preserve the scalar, vector or tensor structure
of the metric perturbation). These variables have simple
physical interpretations only for certain specific gauge choices.
Finally, note that we have not yet had to specify whether the background FRW
model is open, flat or closed. However, we have made the implicit
assumption that the universe is almost FRW when specifying the zero and
first-order variables in the linearisation procedure.

\section{The CMB Temperature Anisotropy}
\label{sec_cmbtemp}

The energy-integrated moments $\Jgaml_{a_{1}\ldots a_{l}}$ of the photon
distribution function provide a fully covariant description of the CMB
temperature anisotropy. In the $u^{a}$ frame, denote the average bolometric
temperature on the sky at the point $x$ by $T_{0}(x)$, so that
\begin{equation}
T_{0}^{4} \propto {\frac{1}{4\pi}} \int dE d\Omega \, E^{3} \fgam(x,p)
= {\frac{1}{4\pi}} \Jgamlin{0},
\end{equation}
which is just the Stefan-Boltzmann law. We use the fractional temperature
variation $\delta_{T}(e)$ from the full-sky average $T_{0}$ to
characterise the temperature perturbation along the spatial direction $e$
in a gauge-invariant and covariant manner (Maartens et al.\ 1995;
Dunsby 1997).
It follows that
\begin{equation}
\bigl( 1+ \delta_{T}(e)\bigr)^{4} = {\frac{4\pi}{\Jgamlin{0}}}
\int dE \, E^{3} \fgam(x,p),
\end{equation}
so that to first-order
\begin{equation}
\delta_{T}(e) = {\frac{1}{4 \rho^{(\gamma)}}} \sum_{l=1}^{\infty}
{\frac{(2l+1)(2l)!}{(-2)^{l} (l!)^{2}}} \Jgaml_{a_{1} \ldots a_{l}}
e^{a_{1}} \ldots e^{a_{l}}.
\label{eq_deltat}
\end{equation}
The right-hand side of equation~\eqref{eq_deltat} is the covariant angular
expansion of the temperature anisotropy.
The tensors $\Jgaml_{a_{1}\ldots a_{l}}$
thus provide a natural covariant description of the CMB anisotropy.
They may be related to the more familiar $a_{lm}$ components in the spherical
harmonic expansion of $\delta_{T}(e)$ by introducing an orthogonal triad
in the instantaneous rest space at $x$, so that $e^{a} = (\snthet\csphi,
\snthet\snphi,\csthet)$. Then the two expansions are related by
\begin{equation}
4 \rho^{(\gamma)} \sum_{m=-l}^{l} a_{lm} Y_{lm}(\theta,\phi) =
{\frac{(2l+1)(2l)!}{(-2)^{l} (l!)^{2}}} \Jgaml_{a_{1} \ldots a_{l}}
e^{a_{1}} \ldots e^{a_{l}}.
\end{equation}
Squaring this expression and integrating over solid angles, we find the
important rotational invariant
\begin{equation}
{\frac{1}{2l+1}} \sum_{m=-l}^{l} |a_{lm}|^{2} =
{\frac{4\pi}{\left(4\rho^{(\gamma)}\right)^{2}}}
\frac{(2l)!}{(-2)^{l} (l!)^{2}}
h^{a_{1}b_{1}} \ldots h^{a_{l}b_{l}} \Jgaml_{a_{1}\ldots a_{l}}
\Jgaml_{b_{1}\ldots b_{l}}.
\label{eq_pow}
\end{equation}
The quantity on the left is a quadratic estimator for the CMB power spectrum
(the $C_{l}$), which we see is related to the covariant tensors
$\Jgaml_{a_{1}\ldots a_{l}}$ in a very simple manner. Further properties
of the covariant and gauge-invariant description of CMB temperature
anisotropies are given by Gebbie \&\ Ellis (1998).

\section{Scalar Perturbations}
\label{sec_scal}

Up to this point, we have treated the scalar, vector and tensor modes of linear
theory in a unified manner. However, to obtain solutions to the covariant
equations it proves useful to consider scalar, vector and tensor modes
separately. In this section we consider scalar modes; tensor modes
are discussed briefly in Section~\ref{sec_tens}.
(Vector modes decay in an expanding universe in the absence of defects, and
so are not likely to have a significant effect on the CMB in inflationary
models.) In the covariant approach to perturbations in
cosmology, we characterise scalar perturbations by demanding that the
magnetic part of the Weyl tensor and the vorticity be at most
second-order. Setting $\clb_{ab}=\ord(2)$ ensures that gravitational waves
are excluded to first-order, and demanding that $\varpi_{ab}=\ord(2)$ ensures
that the density gradients seen by an observer in the $u^{a}$ frame arise
from clumping of the density, $\sgrad^{2} \rho =\ord(1)$, and not from
kinematic effects due to vorticity (the absence of flow-orthogonal
hypersurfaces), which give $\sgrad^{2} \rho = \ord(2)$ in an almost-FRW
universe. Note that we do not classify scalar perturbations as having
$\clb_{ab}=\varpi_{ab}=0$ (to all orders), which is only a highly restricted
subset of the full set of scalar solutions. For example, in an (exactly)
irrotational dust-filled universe (a ``silent'' universe), it can be shown
from the exact non-linear equations that demanding $\clb_{ab}=0$ forces the
solution into a very small class, which probably all have high
symmetry~\cite{ellis-er96}, and so cannot represent a very general
perturbation. This arises because requiring that $\clb_{ab}=0$ be preserved
along the flow lines introduces a series of complex constraints which
reduce greatly the size of the solution set.
However, requiring only that $\clb_{ab}$ and $\varpi_{ab}$ be at
most second-order gives a much larger class of solutions because
only two new constraints are introduced, and these are necessarily preserved
by the propagation equations.
The solutions with $\clb_{ab}$ and $\varpi_{ab}$ vanishing exactly comprise
a very small subset of the larger class of exact solutions which we classify as
scalar perturbations.

On setting $\clb_{ab}=0$ and $\varpi_{ab}=0$, (equality to zero in the
linearised theory should be taken to imply that the quantity is at
most second-order) we see from equation~\eqref{eq_cons1} that
\begin{equation}
\sgrad^{c} \sigma_{d(a}{\eta_{b)ce}}^{d} u^{e} = 0 \quad \Rightarrow \quad
\sgrad_{[a} \sgrad^{c} \sigma_{b]c} = 0,
\label{eq_sigscalcons}
\end{equation}
where the antisymmetrisation is on the indices $a$ and $b$ in the right-hand
equation.
This is a necessary condition for $\sigma_{ab}$ to be constructed
from a scalar potential.
It follows from equations~\eqref{eq_cons4} and~\eqref{eq_prop4}
that
\begin{equation}
\sgrad_{[a}q_{b]} = 0,\qquad \sgrad_{[a}w_{b]} = 0,
\end{equation}
so that the heat flux and acceleration may be written as spatial gradients
of scalar fields (making use of the integrability condition given as
equation~\eqref{eq_intcon}).
Consistency of $\sgrad_{[a}q_{b]} = 0$ with
equation~\eqref{eq_prop5} for $q_{a}$ then requires that
\begin{equation}
\sgrad_{[a} \sgrad^{c} \pi_{b]c} = 0 \quad \Rightarrow \quad
\sgrad_{[a} \sgrad^{c} \cle_{b]c} = 0,
\end{equation}
with the implication following from equation~\eqref{eq_cons3}. It follows that
all vector variables, such as $q_{a}$ and $\sgrad^{b} \cle_{ab}$, may
be derived from scalar potentials. The new constraint, given as
equation~\eqref{eq_sigscalcons},
is only consistent with the propagation equations if $\cle_{ab}$ and
$\pi_{ab}$ satisfy
\begin{equation}
\sgrad^{c} \cle_{d(a}{\eta_{b)ce}}^{d} u^{e} = 
-{\tfrac{1}{2}} \kappa \sgrad^{c} \pi_{d(a}{\eta_{b)ce}}^{d} u^{e}.
\label{eq_clescalcons}
\end{equation}
In the absence of anisotropic stress, we see that the left-hand side
of equation~\eqref{eq_clescalcons} is constrained to be zero, which is
consistent with the propagation equation for $\cle_{ab}$,
given as equation~\eqref{eq_prop1},
with $\pi_{ab}=0$. If the anisotropic stress does not vanish, we
include the constraint
\begin{equation}
\sgrad^{c} \cle_{d(a}{\eta_{b)ce}}^{d} u^{e} = 0 \quad \Rightarrow \quad
\sgrad^{c} \pi_{d(a}{\eta_{b)ce}}^{d} u^{e} = 0,
\label{eq_piscalcons}
\end{equation}
in the definition of a scalar mode, which is easily shown to be consistent
with the propagation equation for $\cle_{ab}$. Requiring consistency
of equation~\eqref{eq_piscalcons} with the propagation equation for $\pi_{ab}$
implied by the photon and neutrino Boltzmann hierarchy yields a series
of constraints on the moments $\Jgaml_{a_{1}\ldots a_{l}}$ and $\Gnul_{a_{1}
\ldots a_{l}}$, which are necessary conditions for them to be derived from
scalar potentials.

The new constraint equations that we have introduced, by restricting the
solution to be a scalar mode, may be satisfied by constructing the covariant
and gauge-invariant variables from tensors derived from scalar potentials
by taking appropriate spatial covariant derivatives of the scalar functions.
It proves convenient to separate the temporal and spatial aspects of the
problem by expanding the scalar potentials in the eigenfunctions
$\qk$ of the generalised Helmholtz equation (Hawking 1966;
Ellis et al.\ 1989)
\begin{equation}
\sgrad^{2} \qk \equiv \sgrad^{a} \sgrad_{a} \qk = {\tfrac{k^{2}}{S^{2}}} \qk,
\label{eq_scalqk}
\end{equation}
which are constructed to satisfy
\begin{equation}
\qkdt = 0.
\label{eq_scalqkdt}
\end{equation}
These equations are correct to zero-order only; in such equations, equality
should be understood to this order only.
In general, we cannot impose equation~\eqref{eq_scalqk} and
$\qkdt=0$ exactly, since the constraint equation is inconsistent with the
evolution implied by $\qkdt=0$ at first-order.
The allowed values of the eigenvalues $k^{2}/S^{2}$ are determined by the
scalar curvature of the background model (since $\qk$ are only
needed to zero-order). In a flat model, $K=0$, $k$ is
a comoving continuous wavenumber $\geq 0$. In closed models, $K>0$, $k$ takes
only discrete values with $k^{2} = \gamma (\gamma +2)K$ where $\gamma$ is a
non-zero, positive integer. In open models, $K<0$, $k$ again takes continuous
values, but with the restriction $k^{2} \geq |K|$.
More details may be found in Harrison (1967). The eigenfunctions
$\qk$ are labelled by the lumped index $k$. This index, which determines the
eigenvalue $k^{2}/S^{2}$, should be understood to distinguish implicitly the
distinct degenerate eigenfunctions which all have the same eigenvalue
$k^{2}/S^{2}$. This multiple use of the symbol $k$ should not cause
any confusion, since the lumped index will always appear as a superscript
or subscript. A function of the eigenvalue $k$ will be denoted with the
eigenvalue as an argument, for example $A(k)$, to distinguish it from the
quantity $A_{k}$ which depends on the mode label $k$ and not just the
eigenvalue.
From the $\qk$ we form a vector $\qk_{a}$,
\begin{equation}
\qk_{a} \equiv {\tfrac{S}{k}} \sgrad_{a} \qk,
\end{equation}
which is orthogonal to $u^{a}$ and is parallel transported at zero-order
along the flow lines:
\begin{equation}
u^{a} \qk_{a} = 0 , \qquad \qkdt_{a} = 0.
\end{equation}
We define totally symmetric tensors of rank $l$, $\qk_{a_{1}\ldots a_{l}}$,
by the recursion formula (for $l > 1$) (see also Gebbie \&\ Ellis (1998))
\begin{equation}
\qk_{a_{1}\ldots a_{l}} = {\tfrac{S}{k}} \left(
\sgrad_{(a_{1}} \qk_{a_{2} \ldots a_{l})} - {\tfrac{l-1}{2l-1}}
\sgrad^{b} \qk_{b (a_{1} \ldots a_{l-2}} h_{a_{l-1} a_{l})}\right).
\label{eq_recurs}
\end{equation}
These tensors satisfy the zero-order properties
\begin{equation}
u^{a_{1}} \qk_{a_{1} a_{2} \ldots a_{l}} = 0, \qquad
h^{a_{1} a_{2}} \qk_{a_{1} a_{2} \ldots a_{l}} = 0, \qquad
\qkdt_{a_{1} a_{2} \ldots a_{l}} = 0,
\end{equation}
which are readily proved by induction.

The scalar functions $\qk$ are the covariant generalisations of the
scalar eigenfunctions of the Laplace-Beltrami operator on the homogeneous
spatial sections of the background FRW model, which are usually employed
in the harmonic decomposition of perturbed quantities (see,
for example, Bardeen (1980)). In the covariant approach, attention
is focused on a velocity field $u^{a}$, rather than a spatial slicing
of spacetime, so it is natural to employ harmonic functions defined
by equation~\eqref{eq_scalqk}. Some of the differential properties of the
derived tensors $\qk_{a{1} \ldots a_{l}}$ (for $l \leq 2$) are given in the
appendix to Bruni et al.\ (1992). We add two more results to this list
which will be useful later:
\begin{eqnarray}
\sgrad^{a_{1}} \qk_{a_{1} a_{2} \ldots a_{l}} &=&
{\tfrac{l}{2l-1}} {\tfrac{k}{S}} \left[ 1 - (l^{2}-1) {\tfrac{K}{k^{2}}}
\right] \qk_{a_{2} \ldots a_{l}} \\
\sgrad^{2} \qk_{a_{1} \ldots a_{l}} &=& {\tfrac{k^{2}}{S^{2}}} \left[1-
l(l+1) {\tfrac{K}{k^{2}}} \right] \qk_{a_{1} \ldots a_{l}},
\end{eqnarray}
which may be derived from the recursion relation given as
equation~\eqref{eq_recurs} and the definition in equation~\eqref{eq_scalqk}.
Further properties of the scalar harmonics are given by Gebbie \&\
Ellis (1998).

To relate the timelike integration employed in the Boltzmann multipole
approach, adopted here, to the lightlike integration employed in
line of sight methods~\cite{seljak96}, which we consider in
Section~\ref{sec_int}, it is convenient to note the following zero-order
results for the variation of the $\qk_{a_{1}\ldots a_{l}}$ along the line
of sight through some point $R$.
Let $x^{a}(\lambda)$ be a null geodesic with tangent vector
parallel to $u_{a} + e_{a}$, and $\lambda$ satisfying
$(u_{a}+e_{a})\nabla^{a}\lambda=1$.
Define a positive parameter $y(\lambda)$ along the past null geodesic by
$dy/d\lambda = -k/S$ with $y=0$ at the point of observation $R$.
Then the evolution of the quantities $\qk_{a_{1}\ldots a_{l}} e^{a_{1}}\dots
e^{a_{l}}$ is given to zero-order by the hierarchy
\begin{eqnarray}
\frac{d}{dy} \left(\qk_{a_{1}\ldots a_{l}} e^{a_{1}}\dots e^{a_{l}}\right)
&=& - \qk_{a_{1}\ldots a_{l+1}} e^{a_{1}}\dots e^{a_{l+1}} \nonumber\\
&&\mbox{}+\frac{l^{2}}{(4l^{2}-1)} \left[1-\frac{K}{k^{2}}(l^{2}-1)\right]
\qk_{a_{1}\ldots a_{l-1}} e^{a_{1}}\dots e^{a_{l-1}},
\label{eq_qloshier}
\end{eqnarray}
which follows from the recursion relation given as equation~\eqref{eq_recurs}.
We shall only consider the solution to this hierarchy in a $K=0$ universe,
in which case we find the following variation of $\qk$ along the line
of sight:
\begin{equation}
(\qk)_{y} = \sum_{l=0}^{\infty} \frac{(2l)!(2l+1)}{(-2)^{l}(l!)^{2}}
j_{l}(y)(\qk_{a_{1}\ldots a_{l}} e^{a_{1}}\dots e^{a_{l}} )_{y=0},
\label{eq_qlos}
\end{equation}
where the $j_{l}(y)$ are spherical Bessel functions. Equation~\eqref{eq_qlos}
expresses $\qk$ a parameter distance $y$ down the line of sight in terms of
the $\qk_{a_{1}\ldots a_{l}} e^{a_{1}}\dots e^{a_{l}}$ at the point of
observation, $R$. If required, the value of $\qk_{a_{1}\ldots a_{l}}
e^{a_{1}}\dots e^{a_{l}}$ down the lightcone can be found from the
solution for $\qk$ (eq.~\eqref{eq_qlos}) and the hierarchy
(eq.~\eqref{eq_qloshier}).

The additional constraints introduced by the conditions for a scalar mode
are satisfied identically if we construct the gauge-invariant variables
in the following manner:
\begin{eqnarray}
\clx_{a}^{(i)} &=& \sum_{k} k \clx_{k}^{(i)} \qka \\
q^{(i)}_{a} &=& \rho^{(i)} \sum_{k} q_{k}^{(i)} \qka \\
v^{(i)}_{a} &=& \sum_{k} v_{k}^{(i)} \qka \\
\pi^{(i)}_{ab} &=& \rho^{(i)} \sum_{k} \pi_{k}^{(i)} \qkab \label{eq_qpi}\\
\clz_{a} &=& \sum_{k} {\tfrac{k^{2}}{S}} \clz_{k} \qka \\
\cle_{ab} &=& \sum_{k} {\tfrac{k^{2}}{S^{2}}} \Phi_{k} \qkab \\
\sigma_{ab} &=& \sum_{k} {\tfrac{k}{S}} \sigma_{k} \qkab \\
w_{a} &=& \sum_{k} {\tfrac{k}{S}} w_{k} \qka,
\end{eqnarray}
where $i$ labels the particle species (and we omit the label when referring
to total fluid variables). The symbolic summation in these expressions is
a sum over eigenfunctions of equation~\eqref{eq_scalqk}. For closed background
models, the sum is discrete, but in the flat and open cases the summation
should be understood as an integral over the continuous label $k$, which
distinguishes distinct eigenfunctions.
The scalar expansion coefficients, such as
$\clx^{(i)}_{k}$, are themselves first-order gauge-invariant variables,
which satisfy
\begin{equation}
\sgrad^{a} \clx^{(i)}_{k} = \ord(2).
\end{equation}
They are labelled by the lumped index $k$.
Finally, we assume that the higher-order angular moments of the photon
and neutrino distribution functions may also be expanded in the
$\qk_{a_{1} \ldots a_{l}}$ harmonics. By considering the zero-order form of
the scalar harmonics $\qk$, and derived tensors, it is straightforward to
show that this condition is equivalent to the usual assumption that the
Fourier components of the distribution functions are axisymmetric about
the wavevector $\bk$ (see, for example, Seljak (1996)). With this
condition, we have
\begin{equation}
\Jgaml_{a_{1}\ldots a_{l}} = \rho^{(\gamma)} \sum_{k} \Jgamlk \qk_{a_{1}\ldots
a_{l}}, \qquad
\Gnul_{a_{1}\ldots a_{l}} = \rho^{(\nu)} \sum_{k} \Gnulk \qk_{a_{1}\ldots
a_{l}},
\label{eq_gamharm}
\end{equation}
for photons and neutrinos respectively.

\subsection{The Scalar Equations}
\label{sec_scal_eqs}

It is now a simple matter to substitute the harmonic expansions of the
covariant variables into the constraint and propagation equations given
in Sections~\ref{sec_cov} and~\ref{sec_eqs}, to obtain equations for the
scalar expansion coefficients that describe scalar perturbations in a
covariant and gauge-invariant manner. To simplify matters, we assume that
the variations in baryon pressure $p^{(b)}$ due to entropy variations are
negligible compared to those arising from variations in $\rho^{(b)}$,
so that we may write
\begin{equation}
\nabla^{a} p^{(b)} = \csound^{2} \, \nabla^{a} \rho^{(b)},
\end{equation}
where $\csound$ is the adiabatic sound speed in the baryon/electron fluid
(this is different to the total sound speed in the tightly-coupled
baryon/photon fluid).

With this assumption, we obtain the following equations for scalar
perturbations: for the spatial gradients of the densities, we find
\begin{eqnarray}
\dot{\clx}^{(\gamma)}_{k} &=& -{\tfrac{k}{S}}\left(
{\tfrac{4}{3}} \clz_{k} + q^{(\gamma)}_{k} \right)+
{\tfrac{4}{3}}\theta w_{k} \label{eq_sec_scal_1}\\
\dot{\clx}^{(\nu)}_{k} &=& -{\tfrac{k}{S}}\left(
{\tfrac{4}{3}} \clz_{k} + q^{(\nu)}_{k} \right)+
{\tfrac{4}{3}}\theta w_{k} \\
\dot{\clx}^{(c)}_{k} &=& -{\tfrac{k}{S}}\left(\clz_{k} + v^{(c)}_{k}\right)
+\theta w_{k} \\
\dot{\clx}^{(b)}_{k} &=& \left(1 + {\tfrac{p^{(b)}}{\rho^{(b)}}} \right)
\left[-{\tfrac{k}{S}}  (\clz_{k} + v^{(b)}_{k}) + \theta w_{k}
\right] + \left({\tfrac{p^{(b)}}{\rho^{(b)}}} - \csound^{2}\right)
\theta \clx^{(b)}_{k}\label{eq_sec_scal_2},
\end{eqnarray}
for the spatial gradient of the expansion, we find
\begin{eqnarray}
\dot{\clz}_{k} &=& -{\tfrac{1}{3}}\theta \clz_{k} -
{\tfrac{1}{2}}{\tfrac{S}{k}} \kappa\left[ 2 \left(\rho^{(\gamma)}
\clx^{(\gamma)}_{k} + \rho^{(\nu)}\clx^{(\nu)}_{k}\right)
+ \rho^{(c)}\clx^{(c)}_{k} + (1 + 3\csound^{2})\rho^{(b)}\clx^{(b)}_{k}\right]
\nonumber \\
&&\mbox{} + {\tfrac{S}{k}} w_{k} \left[ {\tfrac{3}{2}} \kappa \rho \gamma
- {\tfrac{3K}{S^{2}}} \right],
\end{eqnarray}
where $\gamma$ is defined in terms of the total pressure $p$ and density
$\rho$ by $p=(\gamma-1)\rho$ (note that we do not assume that
$\gamma$ is constant). The heat fluxes satisfy
\begin{eqnarray}
\dot{q}^{(\gamma)}_{k} + {\tfrac{1}{3}}{\tfrac{k}{S}}(2 \pi^{(\gamma)}_{k}
- \clx^{(\gamma)}_{k} + 4 w_{k})
&=& \nelec \sigma_{T} \left({\tfrac{4}{3}} v^{(b)}_{k} - q^{(\gamma)}_{k}
\right) \\
\dot{q}^{(\nu)}_{k} + {\tfrac{1}{3}}{\tfrac{k}{S}}(2\pi^{(\nu)}_{k}
- \clx^{(\nu)}_{k} + 4 w_{k}) &=& 0,
\end{eqnarray}
and for the baryon and CDM peculiar velocities
\begin{eqnarray}
\dot{v}^{(c)}_{k} + {\tfrac{1}{3}} \theta v^{(c)}_{k} + {\tfrac{k}{S}}
w_{k} &=& 0\\
\left(1 + {\tfrac{p^{(b)}}{\rho^{(b)}}} \right) \left[
\dot{v}^{(b)}_{k} + {\tfrac{1}{3}} (1-3\csound^{2}) \theta v^{(b)}_{k}
+{\tfrac{k}{S}} w_{k} \right] - {\tfrac{k}{S}} \csound^{2} \clx^{(b)}_{k} &=&
- \nelec \sigma_{T} {\tfrac{\rho^{(\gamma)}}{\rho^{(b)}}}
\left({\tfrac{4}{3}} v^{(b)}_{k} - q^{(\gamma)}_{k}\right).
\end{eqnarray}
The propagation equations for the anisotropic stresses are
\begin{eqnarray}
\dot{\pi}^{(\gamma)}_{k} + {\tfrac{3}{5}}{\tfrac{k}{S}} \left(1-
{\tfrac{8K}{k^{2}}}\right) \Jgamlink{3} - {\tfrac{2}{5}}{\tfrac{k}{S}}
q^{(\gamma)}_{k} - {\tfrac{8}{15}}{\tfrac{k}{S}} \sigma_{k} &=&
-{\tfrac{9}{10}} \nelec \sigma_{T} \pi^{(\gamma)}_{k} \\
\dot{\pi}^{(\nu)}_{k} + {\tfrac{3}{5}}{\tfrac{k}{S}} \left(1-
{\tfrac{8K}{k^{2}}}\right) \Gnulink{3} - {\tfrac{2}{5}}{\tfrac{k}{S}}
q^{(\nu)}_{k} - {\tfrac{8}{15}}{\tfrac{k}{S}} \sigma_{k} &=& 0,
\end{eqnarray}
and the remaining moment equations, for $l \geq 3$, are
\begin{eqnarray}
\dot{J}^{(l)}_{k} + {\tfrac{k}{S}}\left\{ {\tfrac{l+1}{2l+1}}
\left[1-l(l+2) {\tfrac{K}{k^{2}}}\right] \Jgamlink{l+1} -
{\tfrac{l}{2l+1}} \Jgamlink{l-1} \right\} &=& - \nelec \sigma_{T} \Jgamlk
\\
\dot{G}^{(l)}_{k} + {\tfrac{k}{S}}\left\{ {\tfrac{l+1}{2l+1}}
\left[1-l(l+2) {\tfrac{K}{k^{2}}}\right] \Gnulink{l+1} -
{\tfrac{l}{2l+1}} \Gnulink{l-1} \right\} &=& 0.
\end{eqnarray}
The propagation equations for $\cle_{ab}$ and $\sigma_{ab}$ become
\begin{eqnarray}
\left({\tfrac{k}{S}}\right)^{2}(\dot{\Phi}_{k} + {\tfrac{1}{3}}\theta \Phi_{k})
+ \half {\tfrac{k}{S}} \kappa\rho (\gamma \sigma_{k} + q_{k}) +
{\tfrac{1}{6}} \kappa \rho \theta (3\gamma -1) \pi_{k} - \half \kappa \rho
\dot{\pi}_{k} &=& 0
\\
{\tfrac{k}{S}} (\dot{\sigma}_{k} + {\tfrac{1}{3}}\theta \sigma_{k})
+ \left({\tfrac{k}{S}}\right)^{2} (\Phi_{k}-w_{k}) + \half \kappa \rho \pi_{k}
&=& 0.
\end{eqnarray}
Finally, the remaining constraint equations become
\begin{eqnarray}
2 \left({\tfrac{k}{S}}\right)^{3}\left(1-{\tfrac{3K}{k^{2}}}\right) \Phi_{k}
- {\tfrac{k}{S}} \kappa \rho \left[\clx_{k} +
\left(1-{\tfrac{3K}{k^{2}}}\right)\pi_{k}\right] - \kappa \rho \theta q_{k}
&=& 0\\
{\tfrac{2}{3}} \left({\tfrac{k}{S}}\right)^{2} \left[\clz_{k}
- \left(1-{\tfrac{3K}{k^{2}}}\right)\sigma_{k}\right]
+ \kappa \rho q_{k} &=& 0.
\end{eqnarray}
The variables $\clx_{k}$, $q_{k}$ and $\pi_{k}$ refer to the total matter,
and are given in terms of the component variables by
\begin{eqnarray}
\rho \clx_{k} &=& \rho^{(\gamma)} \clx^{(\gamma)}_{k} + \rho^{(\nu)}
\clx^{(\nu)}_{k} + \rho^{(b)} \clx^{(b)}_{k} + \rho^{(c)} \clx^{(c)}_{k} \\
\rho q_{k} &=& \rho^{(\gamma)} q^{(\gamma)}_{k} + \rho^{(\nu)}
q^{(\nu)}_{k} + (\rho^{(b)}+p^{(b)}) v^{(b)}_{k} + \rho^{(c)} v^{(c)}_{k} \\
\rho \pi_{k} &=& \rho^{(\gamma)} \pi^{(\gamma)}_{k} + \rho^{(\nu)}
\pi^{(\nu)}_{k}.
\end{eqnarray}

These equations give a complete description of the evolution of inhomogeneity
and anisotropy from scalar perturbations in an almost-FRW universe with
any spatial curvature. The system closes up once a choice for the velocity
$u^{a}$ is made, and it is straightforward to check that the
constraint equations are consistent with the propagation equations.
The equations for $\Jgamlk$ and $\Gnulk$ for $l \geq 3$ are equivalent
to those usually found in the literature (see, for example,
Ma \&\ Bertschinger (1995) and set $K=0$).
This is because the moments of the perturbed distribution
function, used in such gauge-dependent calculations, are gauge-invariant for
$l\geq 1$. (The $l=1$ moment does depend on the choice of coordinates
in the real universe, but is independent of the mapping onto the background
model, since the background distribution function has no angular dependence.)
Gauge-invariant versions of the usual synchronous-gauge equations~\cite{ma95}
are obtained by taking $u^{a}$ to coincide with the CDM velocity, so that
$w_{a}$ and $v^{(c)}_{a}$ vanish.

\subsection{The Integral Solution}
\label{sec_int}

The integral solution to the Boltzmann multipole equations is central to the
line of sight integration method for the calculation CMB
anisotropies. This method, which has been implemented in the CMBFAST code of
Seljak \&\ Zaldarriaga (1996), provides a very fast route to calculating
the CMB power spectrum. Although we do not make use of this method
for the numerical calculations presented in this paper, it may be
useful to some readers to have available the integral solution to the
covariant and gauge-invariant Boltzmann hierarchy, not least because it
provides the link between the lightlike integrations along the observational
null cone, employed in line of sight methods (which includes the
original calculation by Sachs \&\ Wolfe (1967)), and the timelike
integrations along the flow lines of the velocity field $u^{a}$, employed
in the Boltzmann multipole approach. For simplicity, we restrict
attention to $K=0$ almost-FRW universes.

The (linearised) integral solution to the hierarchy of Boltzmann multipole
equations given in the previous subsection is, for $l\geq1$,
\begin{eqnarray}
\left(\Jgamlk\right)_{R} &=& 4 \int_{0}^{t_{R}} \et{-\tilde{\tau}} \Bigl\{
\left({\tfrac{k}{S}} \sigma_{k} + {\tfrac{3}{16}}\nelec \sigma_{T}
\pi^{(\gamma)}_{k} \right)\left[{\tfrac{1}{3}} j_{l}(\tilde{y})
+ {\tfrac{d^{2}}{d\tilde{y}^{2}}} j_{l}(\tilde{y}) \right]
- \left({\tfrac{k}{S}} w_{k} - \nelec\sigma_{T} v_{k}^{(b)}\right)
{\tfrac{d}{d\tilde{y}}}j_{l}(\tilde{y})
\nonumber\\&&\mbox{}
- \left[{\tfrac{1}{3}}\left({\tfrac{k}{S}} \clz_{k} - \theta w_{k}\right)
- {\tfrac{1}{4}}\nelec \sigma_{T} \clx_{k}^{(\gamma)} \right]
j_{l}(\tilde{y}) \Bigr\} \, d t',
\label{eq_intsol}
\end{eqnarray}
where $(\ )_{R}$ denotes the quantity evaluated at $R$, and
the integral is taken along the flow line of $u^{a}$ through the
point $R$.
Here, $t$ is proper time along the flow line (with $t=t_{R}$ at $R$),
$\tilde{y}=\tilde{y}(t_{R},t')$
is $k$ times the conformal time difference along the flow line between
$t'$ and $t_{R}$: $\tilde{y} \equiv \int_{t'}^{t_{R}} k/S \, dt$,
and $\tilde{\tau}=\tilde{\tau}(t_{R},t')$
is an ``optical depth'' along the flow line:
$\tilde{\tau} \equiv \int_{t'}^{t_{R}} \nelec \sigma_{T} \, dt$. In deriving
equation~\eqref{eq_intsol} we have neglected any contribution from initial
conditions, since these are exponentially suppressed by a factor
$\exp[-\tilde{\tau}(t_{R},0)]$. It is straightforward to verify by
differentiating with respect to $t_{R}$ that equation~\eqref{eq_intsol}
is the solution to the Boltzmann
hierarchy for scalar perturbations in a $K=0$ almost-FRW universe.
Verification for the $l=1$ moment requires the following formal
solution for $\clx_{k}^{(\gamma)}$:
\begin{eqnarray}
\left(\clx_{k}^{(\gamma)}\right)_{R}
&=& 4 \int_{0}^{t_{R}}\et{-\tilde{\tau}} \Bigl\{
\left({\tfrac{k}{S}} \sigma_{k} + {\tfrac{3}{16}}\nelec \sigma_{T}
\pi^{(\gamma)}_{k} \right)\left[{\tfrac{1}{3}} j_{0}(\tilde{y})
+ {\tfrac{d^{2}}{d\tilde{y}^{2}}} j_{0}(\tilde{y}) \right]
- \left({\tfrac{k}{S}} w_{k} - \nelec\sigma_{T} v_{k}^{(b)}\right)
{\tfrac{d}{d\tilde{y}}}j_{0}(\tilde{y})
\nonumber\\&&\mbox{}
- \left[{\tfrac{1}{3}}\left({\tfrac{k}{S}} \clz_{k} - \theta w_{k}\right)
- {\tfrac{1}{4}}\nelec \sigma_{T} \clx_{k}^{(\gamma)} \right]
j_{0}(\tilde{y}) \Bigr\} \, d t',
\label{eq_intsolclx}
\end{eqnarray} 
where again we have neglected the exponentially suppressed contribution from
the initial conditions. In numerical applications, it is convenient to
manipulate equation~\eqref{eq_intsol} further by integrating by
parts~\cite{seljak96}.

The integral in equation~\eqref{eq_intsol} is taken along the flow line
of $u^{a}$ through $R$. However, in the linearised calculation considered
here, the integral can be performed along (any) null geodesic through
$R$ also. To see this, regard $\tilde{y}$ and $\tilde{\tau}$ as the
restrictions
to the flow line of (zero-order) fields in the past lightcone of $R$, with
$\sgrad_{a} \tilde{y} = \ord(1)$ and similarly for $\tilde{\tau}$. Replacing
the measure $dt'$ by $u_{a}dx^{a}$ in the integral in
equation~\eqref{eq_intsol}, and noting that $\nabla_{[a}u_{b]}=\ord(1)$
and, for example, $\sgrad_{a}\sigma_{k}=\ord(2)$, it follows that the
line integral of $u^{a}$ times the integrand in equation~\eqref{eq_intsol}
is path-independent. The flow line and null geodesic through $R$ can be joined
at early times by a spacelike curve with $u_{a}dx^{a}=\ord(1)$, which therefore
makes only a second-order contribution to the integral around the closed loop.
To zero-order, the restriction of the fields $\tilde{y}$ and $\tilde{\tau}$
to the null geodesic through $R$ define quantities
$y=y(\lambda_{R},\lambda')$ and $\tau=\tau(\lambda_{R},\lambda')$ on the null
curve, where $y\equiv \int_{\lambda'}^{\lambda_{R}} k/S \, d\lambda$ and
$\tau\equiv \int_{\lambda'}^{\lambda_{R}} \nelec\sigma_{T} \,
d\lambda$, so that $\tau$ is the optical depth along the line of sight.
(The parameter $\lambda$ along the null geodesic was defined in
Section~\ref{sec_scal}.) Using the integral along the line of sight,
we can now reassemble the gauge-invariant
temperature perturbations from the mean, $\delta_{T}(e)$, at $R$ using
equations~\eqref{eq_deltat} and~\eqref{eq_gamharm}. Recalling
equation~\eqref{eq_qlos} for the variation of the quantities
$\qk_{a_{1}\ldots a_{l}} e^{a_{1}}\dots e^{a_{l}}$ down the line of
sight, the temperature anisotropy from scalar perturbations in
an almost-FRW universe reduces to
\begin{eqnarray}
\left(\delta_{T}(e)\right)_{R} &=& - {\tfrac{1}{4}} \sum_{k}
\left(\clx^{(\gamma)}_{k} \qk\right)_{R} \nonumber \\
&&\mbox{} \sum_{k} \int^{\lambda_{R}} \et{-\tau} \left[
{\tfrac{k}{S}} \sigma_{k} \qk_{ab}e^{a}e^{b} + {\tfrac{k}{S}} w_{k}
\qk_{a}e^{a}
- {\tfrac{1}{3}}\left( {\tfrac{k}{S}} \clz_{k} - \theta w_{k}\right)\qk
\right] \, d\lambda' \nonumber \\
&&\mbox{} \sum_{k} \int^{\lambda_{R}} \nelec \sigma_{T} \et{-\tau}
\left[{\tfrac{3}{16}}\pi_{k}^{(\gamma)} \qk_{ab}e^{a}e^{b} - v_{k}^{(b)}
\qk_{a}e^{a} + {\tfrac{1}{4}} \clx_{k}^{(\gamma)} \qk \right] \, d\lambda'.
\label{eq_intsolK}
\end{eqnarray}
Equation~\eqref{eq_intsolK}, first given in this covariant form in
Challinor \&\ Lasenby (1998),
is valid for any value of the spatial curvature, even though the
derivation given here considers the $K=0$ case only.
Equation~\eqref{eq_intsolK} is most easily obtained by a direct integration
of the covariant Boltzmann equation for the temperature anisotropy
$\delta_{T}(e)$ along the line of sight~\cite{LC-sw}. However, the route
followed here makes clear the link between the lightlike and
timelike integrations, employed in the line of sight and
Boltzmann multipole methods respectively.

\subsection{Initial Conditions on Super-Horizon Scales}

In this subsection, we analytically extract the solution of the scalar
perturbation equations in the radiation dominated era. We shall only
consider modes with $|K|/k^{2} \ll 1$ so that we may ignore terms involving
$K$ in the scalar equations. Associated with each mode there is a
characteristic length scale, $S/k$. The condition $|K|/k^{2}\ll 1$ is
equivalent to requiring that this length scale be small
compared to the curvature radius of the universe. For such modes, $k$
is effectively a comoving wavenumber. We shall also require that the
mode be well outside the horizon scale $1/H$, so that we consider only those
modes satisfying
\begin{equation}
1 \ll \clh_{k}^{2} \ll {\tfrac{H^{2}S^{2}}{|K|}},
\label{eq_modecon}
\end{equation}
where $\clh_{k} \equiv S H /k$ is the ratio of the characteristic length scale
to the horizon scale, and $H^{2}S^{2}/|K|$ is the (squared) ratio
of the curvature radius to the horizon scale. If the universe may be
approximated by a $K=0$ universe to zero-order, equation~\eqref{eq_modecon}
reduces to $\clh_{k} \gg 1$. The approximate analytic
solution may be used to provide initial conditions for a numerical integration
of the scalar equations (see Section~\ref{sec_num}).

At this point, it is convenient to make a choice of frame. In the CDM model,
the rest frame of the CDM defines a geodesic frame, which provides
a convenient choice for $u^{a}$, since the acceleration then vanishes
identically. We assume that this frame choice has been made in the
rest of this paper.

Well before decoupling, the baryons and photons are tightly-coupled
because of the high opacity to Thomson scattering. This scattering damps
the photon moments for $l\geq 2$, but a dipole ($l=1$)
moment can survive if the
baryon velocity does not coincide with the CDM velocity. To a good
approximation, we may ignore the $\Jgamlk$ for $l\geq 2$, and set
$v^{(b)}_{k} = 3 q^{(\gamma)}_{k}/4$ so that the radiation is isotropic
in the rest frame of the baryons. This is the lowest-order term in the
tight-coupling approximation (see Section~\ref{sec_tightcoup}).
Similarly, we expect that the higher-order neutrino moments
will also be small in the early universe, since the neutrinos were in thermal
equilibrium prior to their decoupling. Furthermore, the baryon and CDM
densities, $\rho^{(b)}$ and $\rho^{(c)}$, are negligible compared to the
radiation and neutrino densities, $\rho^{(\gamma)}$ and $\rho^{(\nu)}$,
in the radiation dominated era.

A useful first approximation to the full set of scalar equations is obtained by
setting the neutrino moments, $\Gnulk$, to zero for $l\geq 2$. It is
convenient to take the dependent variable to be $x\equiv \clh_{k}^{-1}$
instead of the proper time $t$ along the flow lines, so that the scalar
propagation equations of the previous section reduce to the following set:
\begin{eqnarray}
x^{2} \clz_{k}' + x \clz_{k} + 3 \left[(1-R)\clx^{(\gamma)}_{k} +
R \clx^{(\nu)}_{k} \right] &=& 0 \label{eq_eu1}\\
x^{2} \Phi_{k}' + x\Phi_{k} + 2 \sigma_{k} + {\tfrac{3}{2}}
\left[(1-R)q^{(\gamma)}_{k} + R q^{(\nu)}_{k} \right] &=& 0\\
x \sigma_{k} ' + \sigma_{k} + x\Phi_{k} &=& 0\\
\clx^{(\gamma)\prime}_{k} + {\tfrac{4}{3}}\clz_{k} + q^{(\gamma)}_{k} &=& 0\\
\clx^{(\nu)\prime}_{k} + {\tfrac{4}{3}}\clz_{k} + q^{(\nu)}_{k} &=& 0\\
q^{(\gamma)\prime}_{k} - {\tfrac{1}{3}} \clx^{(\gamma)}_{k} &=& 0 \\
q^{(\nu)\prime}_{k} - {\tfrac{1}{3}} \clx^{(\nu)}_{k} &=& 0,
\end{eqnarray}
where a prime denotes differentiation with respect to $x$, and we
have used the zero-order Friedmann equation in the form
$H^{2} = \kappa \rho/3$ since the curvature term may be neglected by
equation~\eqref{eq_modecon}.
We have followed Ma \&\ Bertschinger (1995) by introducing the dimensionless
quantity $R$ defined by
\begin{equation}
R \equiv {\frac{\rho^{(\nu)}}{\rho^{(\nu)}+ \rho^{(\gamma)}}}.
\end{equation}
After neutrino decoupling, $R$ is a constant which depends only on the number
of neutrino species. The remaining equations that we require are the two scalar
constraints which reduce to
\begin{eqnarray}
2 x^{3} \Phi_{k} - 3 x\left[(1-R)\clx^{(\gamma)}_{k} + R \clx^{(\nu)}_{k}
\right] - 9\left[(1-R) q^{(\gamma)}_{k} + R q^{(\nu)}_{k} \right] &=& 0\\
2x^{2} (\clz_{k} - \sigma_{k} ) + 9 \left[(1-R)q^{(\gamma)}_{k} + R
q^{(\nu)}_{k}\right] &=& 0. \label{eq_eu9}
\end{eqnarray}

This set of equations gives a closed equation for $\Phi_{k}$:
\begin{equation}
3x \Phi_{k}'' + 12 \Phi_{k}' + x \Phi_{k} = 0.
\label{eq_phipropeu}
\end{equation}
This equation should be compared to the fourth-order equation for the metric
perturbation variable in the synchronous gauge (see, for example,
Ma \&\ Bershinger (1995)).
The fourth-order equation admits four linearly-independent solutions, but two
of the solutions are gauge modes that arise from mapping an exact FRW universe
to itself. The gauge-invariant approach adopted here ensures that such gauge
modes do not arise. This is evident from equation~\eqref{eq_phipropeu} which
is only a second-order equation. The two linearly-independent solutions
of this equation both describe physical perturbations in the Weyl tensor, which
vanishes for an exact FRW universe.
It is now straightforward to find the general solution of
equations~\eqref{eq_eu1}--\eqref{eq_eu9}. There are two solutions with
non-vanishing Weyl tensor ($\Phi_{k} \neq 0$), which we write as
\begin{eqnarray}
\Phi_{k} &=& -3 y^{-3} [(Cy-D)\cos\!y - (C+Dy) \sin\!y ]
\label{eq_eu20}\\
\clz_{k} &=& 3\sqrt{3} y^{-3} [2(C+Dy)\cos\!y + 2(Cy - D) \sin\!y -
C(2 + y^{2}) ]\\
\sigma_{k} &=& 3\sqrt{3} y^{-2} [D \cos\!y + C \sin\!y - C y] \\
q^{(\gamma)}_{k} &=& - 4\sqrt{3} y^{-1} [C \cos\!y - D \sin\!y - C]
\\
q^{(\nu)}_{k} &=& -{\tfrac{2\sqrt{3}}{R}}y^{-1} [(2RC + Dy)\cos\!y
+ (Cy - 2RD) \sin\!y - 2RC ] \\
\clx^{(\gamma)}_{k} &=& 12 y^{-2} [ (C+Dy) \cos\!y + (Cy-D) \sin\!y - C] \\
\clx^{(\nu)}_{k} &=& {\tfrac{6}{R}} y^{-2} [(2RC-Cy^{2} + 2RDy) \cos\!y
+ (2RCy -2 RD + D y^{2}) \sin\!y - 2 RC ],
\label{eq_eu21}
\end{eqnarray}
where $y \equiv x /\sqrt{3}$, and $C$ and $D$ are constants. There are also
three solutions with vanishing Weyl tensor ($\Phi_{k} = 0$), which we write
as
\begin{eqnarray}
\clz_{k} &=& {\tfrac{\sqrt{3}}{4}} A_{3} y^{-3} \left( 2 + y^{2}\right)
\label{eq_eu10} \\
\sigma_{k} &=& {\tfrac{\sqrt{3}}{4}} A_{3} y^{-1}\\
q^{(\gamma)}_{k} &=& -{\tfrac{1}{\sqrt3}}\left(A_{1} \cos\!y
- A_{2} \sin\!y + A_{3} y^{-1} \right) \\
q^{(\nu)}_{k} &=& {\tfrac{R-1}{\sqrt{3}R}} \left(-A_{1} \cos\!y + A_{2}\sin\!y
\right) - {\tfrac{1}{\sqrt{3}}} A_{3} y^{-1} \\
\clx^{(\gamma)}_{k} &=& A_{1} \sin\!y + A_{2} \cos\!y + A_{3} y^{-2} \\
\clx^{(\nu)}_{k} &=& \tfrac{R-1}{R} \left( A_{1} \sin\!y + A_{2} \cos\!y\right)
+ A_{3} y^{-2},\label{eq_eu11}
\end{eqnarray}
where $A_{1}$, $A_{2}$ and $A_{3}$ are further constants. The solution
with only $A_{3}$ non-zero describes a radiation dominated universe which is
exactly FRW except that the CDM has a peculiar velocity (relative to the
velocity of the fundamental observers) $v^{(c)}_{a} = v^{(0)}_{a} / S$, where
$v^{(0)}_{a}$ is a first-order vector, orthogonal to the fundamental
velocity, which is parallel transported along the fundamental flow lines:
$\dot{v}^{(0)}_{a} = \ord(2)$. This can be seen most clearly by adopting the
energy frame, defined by the condition $q_{a}=0$. This is arguably a better
choice to make in the early universe, since $u^{a}$ is then defined in terms
of the dominant matter components, rather than a minority component, such as
the CDM, which has little effect on the gravitational dynamics. Choosing the
energy frame, and ignoring anisotropic stresses (which are frame-independent
in linear theory), the CDM relative velocity evolves according to
\begin{equation}
\dot{v}^{(c)}_{a} + {\tfrac{1}{3}}\theta v^{(c)}_{a} + {\tfrac{1}{4S}}
\left[ (1-R) \clx^{(\gamma)}_{a} + R \clx^{(\nu)}_{a} \right] = 0,
\label{eq_eucdm}
\end{equation}
in the radiation dominated era. Since the CDM interacts with the other matter
components through gravity alone, and since the gravitational influence of the
CDM on the dominant matter components may be ignored during radiation
domination, equation~\eqref{eq_eucdm} is the only equation governing the
evolution of the perturbations which makes reference to the CDM. It follows
that any solution of equation~\eqref{eq_eucdm}, defines
a valid solution to the linearised perturbation equations. The solution,
which has $v^{(c)}_{a} = v^{(0)}_{a} / S$ in an otherwise FRW universe
corresponds to
the solution labelled by $A_{3}$ in equations~\eqref{eq_eu10}--\eqref{eq_eu11}.
Note that this solution decays in an expanding universe. Following standard
practice, we assume that this mode may be ignored ($A_{3}=0$) since it would
require highly asymmetric initial conditions at the end of the inflationary
epoch if the decaying mode was significant during the epoch of interest here.
Similar comments apply to the mode labelled by $D$ in
equations~\eqref{eq_eu20}--\eqref{eq_eu21}.

An important subclass of these solutions describe adiabatic modes. We assume
that the appropriate covariant and gauge-invariant definition of adiabaticity
is that
\begin{equation}
\frac{\sgrad_{a} \rho^{(i)}}{\rho^{(i)} + p^{(i)}} =
\frac{\sgrad_{a} \rho^{(j)}}{\rho^{(j)} + p^{(j)}},
\label{eq_eu_adiab}
\end{equation}
for all species $i$ and $j$ (Bruni et al.\ 1992).
This condition, which is frame-independent in linear theory, is the natural
covariant generalisation of the (gauge-invariant) condition
\begin{equation}
\frac{\delta \rho^{(i)}}{\bar{\rho}^{(i)} + \bar{p}^{(i)}} =
\frac{\delta \rho^{(j)}}{\bar{\rho}^{(j)} + \bar{p}^{(j)}},
\end{equation}
where $\delta \rho^{(i)} = \rho^{(i)} - \bar{\rho}^{(i)}$ is the usual
gauge-dependent density perturbation, and overbars denote the background
quantity. Demanding adiabaticity between the photons and the
neutrinos leaves only one free constant of integration, which we take to
be $C$. The remaining constants are $A_{1}=A_{3}=D=0$ and $A_{2} = -6C$.
Note that the constants of integration will depend on the mode label
$k$, in general, so we have $C=C_{k}$.

This adiabatic solution may be developed further by including higher-moments
of the neutrino distribution function, and finding a series expansion of the
(enlarged) system in $x$. To obtain solutions correct to $\ord(x^{3})$, it
is necessary to retain $\pi^{(\nu)}_{k}$ and $\Gnulink{3}$. The
series solution that results is
\begin{eqnarray}
\Phi_{k} &=& C\left[1 - \tfrac{389R+700}{168(2R+25)(R+5)}x^{2}\right]
\nonumber\\
\clx^{(\gamma)}_{k} &=& \tfrac{(4R+15)C}{6(R+5)} x^{2}  \nonumber\\
\clz_{k} &=& -\tfrac{(4R+15)C}{4(R+5)}\left[x - \tfrac{4R+5}{18(4R+15)}x^{3}
\right]\nonumber\\
\clx^{(\nu)}_{k} &=& \tfrac{(4R+15)C}{6(R+5)} x^{2}  \nonumber \\
\sigma_{k} &=& -\tfrac{5C}{2(R+5)}\left[x +
\tfrac{112R^{2} - 16R - 1050}{2520(2R+25)}x^{3}\right]\nonumber\\
q^{(\gamma)}_{k} &=& \tfrac{(4R+15)C}{54(R+5)}x^{3}  \nonumber \\
\pi^{(\nu)}_{k} &=& -\tfrac{2C}{3(R+5)} x^{2} \nonumber \\ 
q^{(\nu)}_{k} &=& \tfrac{(4R+23)C}{54(R+5)} x^{3} \nonumber \\
\Gnulink{3} &=& -\tfrac{5C}{63(R+5)} x^{3} \nonumber \\
\Gnulk &=& \ord(x^{l}) \quad \mbox{for $l > 3$}.
\label{eq_euser}
\end{eqnarray}
Note that on large scales ($x \ll 1$), the harmonic coefficient $\Phi_{k}$
of $\cle_{ab}$ is constant along the flow lines. It follows that on large
scales, we may write $\cle_{ab} = \cle_{ab}^{(0)} /S^{2}$, where
$\dot{\cle}^{(0)}_{ab}=\ord(2)$. The series solution given in
equation~\eqref{eq_euser} is
adiabatic between the photons and neutrinos to $\ord(x^{3})$, but the
adiabaticity is broken by the higher-order terms. This difference in the
dynamic behaviour of radiation and neutrinos is due to their different
kinetic equations; the neutrinos are collisionless which allows higher-order
angular moments in the distribution function to grow, but the radiation
is tightly-coupled to the baryon fluid which prevents the growth of
higher-order moments. The baryon relative velocity $v^{(b)}_{a}$ is
determined by the condition that the radiation be nearly isotropic in
the rest frame of the baryons:
\begin{equation}
v^{(b)}_{k} = {\tfrac{3}{4}} q^{(\gamma)}_{k},
\end{equation}
and the spatial gradients of the baryon and CDM follow from the adiabaticity
condition:
\begin{equation}
\clx^{(b)}_{k} = \clx^{(c)}_{k} = {\tfrac{3}{4}} \clx^{(\gamma)}_{k} =
{\tfrac{3}{4}} \clx^{(\nu)}_{k},
\end{equation}
where we have neglected the small effect of baryon pressure.
The series solution given as equation~\eqref{eq_euser} was used to provide
adiabatic initial conditions for the numerical solution of the perturbation
equations, discussed in the next section.

\section{Adiabatic Scalar Perturbations in a $K=0$ Universe}
\label{sec_num}

In this section, we discuss the calculation of the CMB power spectrum
from initially adiabatic scalar perturbations in an almost-FRW universe
with negligible spatial curvature.
The evolution of anisotropy in the CMB and inhomogeneities in the density
fields, resulting from scalar perturbations, may be found by solving
numerically the equations presented in Section~\ref{sec_scal}, with
initial conditions determined from the analytic solutions of the previous
section. For adiabatic perturbations, the specification of initial conditions
is particularly simple; there is a single function $C_{k}$ of the mode label $k$
to set. This function gives the (constant) amplitude of the harmonic
component of the electric part of the Weyl tensor on super-horizon scales.

\subsection{The CMB Power Spectrum}

The gauge-invariant temperature perturbation from the mean, denoted
by $\delta_{T}(e)$, is given by equation~\eqref{eq_deltat}.
Substituting for the harmonic expansion of the
angular moments $\Jgaml_{a_{1}\ldots a_{l}}$, we find
\begin{equation}
\delta_{T}(e) = {\frac{1}{4}} \sum_{l=1}^{\infty} {\frac{(2l+1)(2l)!}{(-2)^{l}
(l!)^{2}}} \left[ \sum_{k} C_{k}\Tgamlk \qk_{a_{1}\ldots a_{l}} e^{a_{1}}\ldots
e^{a_{l}} \right],
\label{eq_ps1}
\end{equation}
where we have introduced the radiation transfer function $\Tgamlk$, which is
a function of the eigenvalue $k$ only. The transfer function is defined to
be the value of $\Jgamlk$ for the initial condition $C_{k}=1$.
Since the
dynamics of a scalar mode labelled by the index $k$ depends only on the
eigenvalue of the eigenfunction $\qk$, the transfer function is a function
of the eigenvalue $k$ only. For the linearised theory considered here, we have
$\Jgamlk = C_{k} \Tgamlk$.

We have come as far as we can without making
a specific choice for the scalar harmonic functions $\qk$. To proceed, we
introduce an almost-FRW coordinate system~\cite{ellis-er96} as follows.
If the perturbations in the universe are only of scalar type, then the
velocity $u^{a}$ is hypersurface orthogonal, so that we may label the
orthogonal hypersurfaces with a time label $t$. Furthermore, since we have
chosen $u^{a}$ to be the CDM velocity, which is geodesic, the flow
orthogonal hypersurfaces may be labelled unambiguously with proper
time along the flow lines, so that $u^{a}=\nabla^{a}t$.
The orthogonal hypersurfaces depart from
being spaces of constant curvature only at first-order, so we can introduce
comoving spatial coordinates $x^{i}$, in such a way that our (synchronous)
coordinate system is almost-FRW in form. (Latin indices, such as $i$, run from
1 to 3.) It is then straightforward to show
that the functions $\et{i\bk\!\cdot\!\bx}$, where
$\bk\!\cdot\!\bx=k^{i}x^{i}$ and
$k^{i}$ are constants, satisfy the defining equations for the
scalar harmonic functions, equations~\eqref{eq_scalqk}
and~\eqref{eq_scalqkdt}, with
$k^{2}=k^{i}k^{i}$, in an almost FRW universe with negligible spatial
curvature. It follows that we may take
\begin{equation}
\qk = \et{i\bk\!\cdot\!\bx}.
\end{equation}
For the open and flat cases, the appropriate generalisations of the
$\et{i\bk\!\cdot\!\bx}$~\cite{harrison67} should be used for the $\qk$.
Note that the expansion coefficients, such as $\Jgamlk$, depend on the
detailed choice of the scalar harmonics $\qk$, but that covariant
tensors, such as $\sum_{k}\Jgamlk \qk_{a_{1}\ldots a_{l}}$,
are independent of this choice. If vector perturbations are also significant,
we cannot use the velocity $u^{a}$ to define a time coordinate in the
manner described above. Instead, an almost-FRW coordinate system should be
constructed using an irrotational and geodesic velocity field $\hat{u}^{a}$,
which is close to our chosen fundamental velocity $u^{a}$. (One possibility
is to take $\hat{u}^{a} \propto \nabla^{a}\rho$.) Using this
velocity field, almost-FRW coordinates can be constructed by the above
procedure~\cite{ellis-er96}. The resulting $\qk$ will satisfy the defining
(zero-order) properties of the scalar harmonics in the $u^{a}$ frame, since
the relative velocity of $\hat{u}^{a}$ is first-order.

Since $u^{a} e_{a}=0$, and $u^{a} =
\nabla^{a} t$, we can always choose the $x^{i}$ so that at our observation
point $e^{a} = (0, S^{-1} e^{i})$, with
$e^{i} e^{i} =1$ (for example, one can choose the $x^{i}$ so that
$Sx^{i}$ are locally Cartesian coordinates in the constant time hypersurface).
Then it follows that, to zero-order,
\begin{equation}
\qk_{a_{1}\ldots a_{l}} e^{a_{1}} \ldots e^{a_{l}} =
\frac{(2i)^{l} (l!)^{2}}{(2l)!} P_{l}(\mu) \qk,
\label{eq_ps2}
\end{equation}
where $\mu \equiv k^{i} e^{i} /k$, and $P_{l}(\mu)$ are the Legendre
polynomials. This result demonstrates that expanding the angular moments of
the distribution function in the covariant tensors $\qk_{a_{1}\ldots a_{l}}$
is equivalent to the usual Legendre expansion of the Fourier
modes of the distribution function (which are axisymmetric about the wave
vector $\bk$), in an almost-FRW universe, where the spatial curvature may be
neglected.

Following standard practice, we make the assumption that we inhabit one
realisation of a stochastic ensemble of universes, so that the
$C_{k}$ are random variables. (The physical basis
on which this assumption rests is that initial fluctuations were
generated from causal quantum processes in the early universe, such as
during a period of inflation; see for example, Kolb \&\ Turner (1990).)
Given our chosen form for the $\qk$, statistical isotropy of the ensemble
demands that the covariance matrix for the $C_{k}$ takes the following form:
\begin{equation}
\langle C_{k} C_{k'}^{\ast} \rangle = C^{2}(k) \delta_{k k'},
\label{eq_pscov}
\end{equation}
where $C^{2}(k)$ is the primordial power spectrum which is a function of the
eigenvalue $k$. The $\delta_{k k'}$ appearing in equation~\eqref{eq_pscov} is
defined by $\sum_{k} \delta_{k k'} A_{k} = A_{k'}$, where $A_{k}$ is an
arbitrary
function of the mode label $k$. The CMB power spectrum $C_{l}$ is defined
by $C_{l} \equiv \langle |a_{lm}|^{2} \rangle$, where $a_{lm}$ are the
coefficients in the spherical harmonic expansion of the temperature anisotropy
(see Section~\ref{sec_cmbtemp}). Substituting the harmonic expansion of the
$\Jgaml_{a_{1}\ldots a_{l}}$ into equation~\eqref{eq_pow}, and using the
zero-order result
\begin{equation}
h^{a_{1}b_{1}}\ldots h^{a_{l}b_{l}} \qk_{a_{1}\ldots a_{l}}
\qkstar_{a_{1}\ldots a_{l}} = {\frac{(-2)^{l} (l!)^{2}}{(2l)!}},
\end{equation}
which follows from $\qk = \et{i \bk\!\cdot\!\bx}$, we find the familiar expression
for the CMB power spectrum in terms of the transfer functions and the
primordial power:
\begin{equation}
C_{l} = \pi^{2} \int d \ln \!k\, C^{2}(k)
\left[\Tgamlk\right]^{2}.
\end{equation}

We make the standard assumption that on large scales, the primordial power
spectrum may be approximated by a power-law of the form
$C^{2}(k) \propto k^{n_{s}-1}$. Many inflationary models predict
that the scalar index $n_{s}$ will be close to unity~\cite{kolb-univ}.
The case $n_{s}=1$
describes the scale-invariant spectrum. This term arises from considering
the logarithmic power spectrum in Fourier space of the (gauge-dependent)
fractional density perturbation $\delta_{\rho}$ evaluated at horizon
crossing. An analogous result can be found in the covariant and
gauge-invariant approach: we evaluate the logarithmic power spectrum of the
dimensionless scalar
$\Delta \equiv \sgrad^{2}\rho/(\rho H^{2})$ in the energy frame ($q_{a}=0$).
Making use of equation~\eqref{eq_cons3}, with the contribution from
anisotropic stress neglected, and the
frame-invariance of $\sgrad^{a} \cle_{ab}$ in linear theory, we find that
\begin{equation}
\partial_{\,\ln\!k} \langle \Delta^{2} \rangle = {\tfrac{16}{9}} \pi
\clh_{k}^{-8} C^{2}(k).
\label{eq_ps4}
\end{equation}
Note that $\Delta$ only receives a contribution at linear-order from scalar
modes (Ellis et al.\ 1990).
In deriving equation~\eqref{eq_ps4}, which is valid before the modes labelled
by $k$
reenter the Hubble radius, we have assumed that only the fastest growing scalar
mode is significant so that $\Phi_{k}$ is constant before horizon crossing.
For given $k$, the logarithmic power in $\Delta$ evolves in time due to
the presence of $\clh_{k}$ on the right-hand side of equation~\eqref{eq_ps4}.
However, at horizon crossing $\clh_{k}$ falls below some critical value of
order unity which is independent of $k$. It follows that for the scalar index
$n_{s}=1$, the logarithmic power in $\Delta$ at horizon crossing is
independent of scale.

\subsection{The Tight-Coupling Approximation}
\label{sec_tightcoup}

At early times, when the baryons and photons are tightly-coupled,
the radiation is nearly isotropic in the frame of the baryons. In this
limit, it is convenient to replace the propagation equations for the
$\Jgaml_{a_{1}\ldots a_{l}}$ with $l \geq 1$, and for the baryon relative
velocity $v^{(b)}_{a}$, with approximate equations which may be
developed by an expansion in the photon mean
free path $1/\nelec \sigma_{T}$. The approximate equations are simpler to solve
numerically than the exact equations since the former do not include the
large Thomson scattering terms present in the latter.

For scalar perturbations, it is simplest to work directly with the
harmonic expansion coefficients, $\Jgamlk$ and $v^{(b)}_{k}$. The
relevant time-scales in the problem are the photon mean free time
$t_{c} \equiv (\nelec \sigma_{T})^{-1}$, the expansion time-scale
$t_{H} \equiv H^{-1}$, and the light travel time across the wavelength
of the mode under consideration, $t_{k} \equiv S/k$. In the tight-coupling
approximation, we expand in the small dimensionless numbers
$t_{c}/t_{H}$ and $t_{c}/t_{k}$, so that the procedure is valid for
$t_{c} \ll \mbox{min}(t_{H}, t_{k})$. While a mode is outside the horizon,
$\mbox{min}(t_{H}, t_{k})=t_{H}$, whereas $\mbox{min}(t_{H}, t_{k})=t_{k}$
during the acoustic oscillations. In the CDM frame ($w_{a}=0$), the
procedure is similar to that usually employed (Peebles \&\ Yu 1970;
Ma \&\ Bertschinger 1995) in the
synchronous-gauge. We combine the propagation equations given in
Section~\ref{sec_scal_eqs} for the photon moments $\Jgamlk$ ($l\geq 1$)
and the baryon relative velocity $v^{(b)}_{k}$, to get the exact (in linear
theory) equations:
\begin{eqnarray}
(1+R)\dot{v}^{(b)}_{k} &=& -{\tfrac{3}{4}} R \dot{\Delta}_{k}
-t_{H}^{-1} v^{(b)}_{k} + {\tfrac{1}{4}} R t_{k}^{-1} \clx^{(\gamma)}_{k}
+ \csound^{2} t_{k}^{-1} \clx^{(b)}_{k} 
- {\tfrac{1}{2}} R t_{k}^{-1} \pi^{(\gamma)}_{k} \label{eq_tc1}\\
(1+R) \Delta_{k} &=& -t_{c} \left[ \dot{\Delta}_{k}
-{\tfrac{4}{3}} t_{H}^{-1} v^{(b)}_{k}
-{\tfrac{1}{3}} t_{k}^{-1} \clx^{(\gamma)}_{k}
+ {\tfrac{4}{3}} \csound^{2} t_{k}^{-1} \clx^{(b)}_{k}
+ {\tfrac{2}{3}} t_{k}^{-1} \pi^{(\gamma)}_{k}\right] \\
\pi^{(\gamma)}_{k} &=& -{\tfrac{10}{9}} t_{c} \left[
\dot{\pi}^{(\gamma)}_{k} + {\tfrac{3}{5}} t_{k}^{-1} \Jgamlink{3}
-{\tfrac{2}{5}} t_{k}^{-1} \Delta_{k} - {\tfrac{8}{15}} t_{k}^{-1}
(\sigma_{k} + v^{(b)}_{k}) \right] \\
\Jgamlk &=& - t_{c} \left[\Jgamlkdt + t_{k}^{-1} \left(
{\tfrac{l+1}{2l+1}} \Jgamlink{l+1} - {\tfrac{l}{2l+1}} \Jgamlink{l-1}
\right)\right] \qquad l \geq 3,
\end{eqnarray}
where $\Delta_{k}\equiv q^{(\gamma)}_{k} - 4 v^{(b)}_{k} /3$, and for this
section, $R \equiv 4 \rho^{(\gamma)} / 3 \rho^{(b)}$. Iterating these
equations gives the tight-coupling expansions
\begin{eqnarray}
\dot{v}^{(b)}_{k} &=& \dot{v}^{(b)}_{0} + \dot{v}^{(b)}_{1} + \ldots \\
\Delta_{k} &=& \Delta_{1} + \Delta_{2} + \ldots \\
\pi^{(\gamma)}_{k} &=& \pi^{(\gamma)}_{1} + \pi^{(\gamma)}_{2}
+\ldots \\
\Jgamlk &=& \Jgaml_{l-1} + \Jgaml_{l} + \ldots,
\end{eqnarray}
where the subscript on the variables on the right-hand side denotes the
order in the expansion parameter
$\epsilon = \mbox{max}(t_{c}/t_{H},t_{c}/t_{k})$. To avoid cluttering of
indices, we leave the mode label $k$ implicit.
We shall only require the results to first-order in $\epsilon$:
\begin{eqnarray}
\dot{v}^{(b)}_{0}  &=& {\tfrac{1}{1+R}} \left({\tfrac{1}{4}}{\tfrac{k}{S}}
R \clx^{(\gamma)}_{k} + {\tfrac{k}{S}} \csound^{2} \clx^{(b)}_{k}
-H v^{(b)}_{k}\right) \\
\dot{v}^{(b)}_{1}  &=& -{\tfrac{3R^{2}}{2(1+R)^{2}}} H \Delta_{1}
- {\tfrac{R}{2(1+R)}} {\tfrac{k}{S}} \pi^{(\gamma)}_{1} \nonumber
\\
&& \mbox{} - {\tfrac{R}{4(1+R)^{2}}} \left[ 4 {\tfrac{t_{c}}{t_{H}}} (\dot{H}
+2H^{2})H^{-1} v^{(b)}_{k} + {\tfrac{t_{c}}{t_{k}}} \left(
2H \clx^{(\gamma)}_{k} + \dot{\clx}^{(\gamma)}_{k} - 4\csound^{2}
\dot{\clx}^{(b)}_{k} \right) \right] \\
\Delta_{1} &=& {\tfrac{1}{3(1+R)}}\left[{\tfrac{t_{c}}{t_{k}}}\left(
\clx^{(\gamma)}_{k} - 4\csound^{2} \clx^{(b)}_{k} \right) +
4 {\tfrac{t_{c}}{t_{H}}} v^{(b)}_{k} \right]\\
\pi^{(\gamma)}_{1} &=& {\tfrac{16}{27}}{\tfrac{t_{c}}{t_{k}}}
\left(\sigma_{k} + v^{(b)}_{k} \right).
\end{eqnarray}
The propagation equation for $q^{(\gamma)}_{k}$ in the tight-coupling
approximation may be obtained from the exact equation~\eqref{eq_tc1}, with
$\dot{\Delta}_{k}$ replaced by $\dot{q}^{(\gamma)}_{k}-4\dot{v}^{(b)}_{k}/3$,
and $\dot{v}^{(b)}_{k}$ and $\pi^{(\gamma)}_{k}$ replaced by their
tight-coupling expansions.

\subsection{Numerical Results}

We are now in a position to evolve an initial set of perturbations from
early times to the present in an almost-FRW universe with negligible
spatial curvature, and to calculate
the power spectrum of the CMB anisotropies that result. In this section we
present the results of a numerical simulation in the standard CDM model,
with parameters $H_{0}=50 \kmsmpc$, baryon fraction $\Omega_{b}=0.05$,
CDM fraction $\Omega_{c}=0.95$, Helium fraction $0.24$, zero cosmological
constant, and a scale-invariant spectrum of initially adiabatic conditions
($n_{s}=1$).

Our code to solve the covariant and gauge-invariant perturbation
equations in the CDM frame, including the Boltzmann hierarchies for the
photons and neutrinos, was based on the serial COSMICS code
developed by Bertschinger and Bode, and described in
Ma \&\ Bertschinger (1995).
The COSMICS package, including full
documentation, is available at {http://arcturus.mit.edu/cosmics}.
We modified the COSMICS code to solve the covariant equations given in this
paper, for the matter variables and for the spatial gradient of the
expansion, $\clz_{a}$. The shear, which is required to solve the Boltzmann
hierarchies for the photons and neutrinos, was determined from the
equation~\eqref{eq_cons4}. The electric part of the Weyl tensor, $\cle_{ab}$,
could then be determined from equation~\eqref{eq_cons3}. Our
calculations of the zero-order ionisation history of the universe, which fully
include the effects of Helium and Hydrogen recombination, followed
Ma \&\ Bertschinger (1995), as did our truncation schemes for the
photon and neutrino Boltzmann hierarchies. The first-order tight-coupling
approximation was used at sufficiently early times that
$\mbox{max}(t_{c}/t_{H},t_{c}/t_{k}) \ll 1$.

%%%%%%%%%%%%%%%%%%%%%%%%%%%%%%%%%%%%%%%%%%%%%%%%%%%%%%%%%%%%%%%%%%%%%%%%%%
\begin{figure}[p]
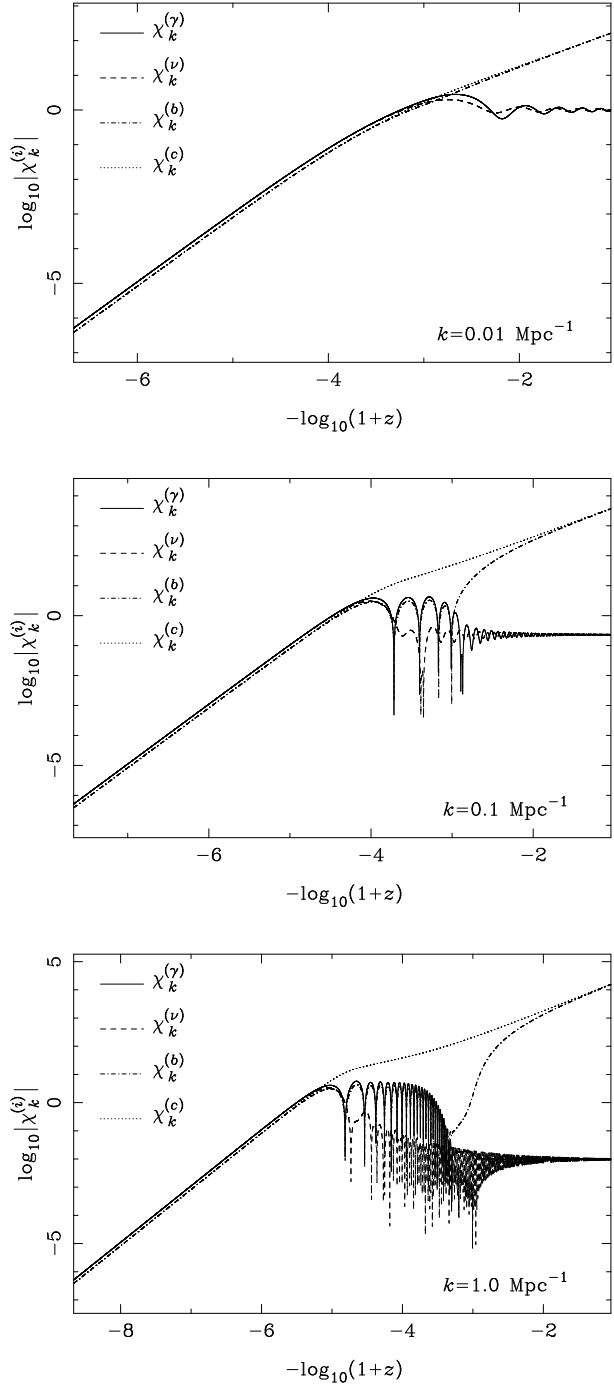

\begin{center}
\begin{picture}(222,525)
\put(-5,525){\hbox{\epsfig{figure=adc98_f1a.ps,angle=-90,width=8cm}}}
\put(-5,345){\hbox{\epsfig{figure=adc98_f1b.ps,angle=-90,width=8cm}}}
\put(-5,165){\hbox{\epsfig{figure=adc98_f1c.ps,angle=-90,width=8cm}}}
\end{picture}
\end{center}
\caption{The variation of the harmonic coefficients
of the fractional comoving density gradients in the CDM frame with redshift in
the standard CDM model: $\Omega_{c}=0.95$, $\Omega_{b}=0.05$,
$\Omega_{\Lambda}=0$, $H_{0}=50\kmsmpc$,
with Helium fraction $0.24$, for $k=0.01$, $0.1$ and $1.0\mpc^{-1}$. The
normalisation is chosen so that $\Phi_{k}=1$ for all $k$ at early times.
\label{fig1}}
\end{figure}
%%%%%%%%%%%%%%%%%%%%%%%%%%%%%%%%%%%%%%%%%%%%%%%%%%%%%%%%%%%%%%%%%%%%%%%%%%

In Figure~\ref{fig1} we show the variation of the harmonic coefficients
$\clx^{(i)}_{k}$ of the comoving fractional spatial gradients in the CDM frame,
against redshift in the standard CDM model. Similar plots were given by
Ma \&\ Bertschinger (1995) for the Fourier components of the
(gauge-dependent) density perturbations $\delta_{\rho}^{(i)} \equiv
(\rho^{(i)} - \bar{\rho}^{(i)} )/ \bar{\rho}^{(i)}$,
where $\bar{\rho}^{(i)}$ is the
density of the species $i$ in the background model. Our results, given in
Figure~\ref{fig1}, agree well with the synchronous-gauge results
of Ma \&\ Bertschinger (1995). This is because the constant time surfaces in
this gauge are orthogonal to the CDM velocity, so that $\clx_{a}^{(i)}$ is a
covariant measure of the density inhomogeneity in these surfaces. Although
$\delta_{\rho}^{(i)}$ is gauge-dependent in the synchronous gauge, the
gauge-conditions restrict this gauge-dependence to transformations of the form
$\delta_{\rho}^{(i)} \mapsto \delta_{\rho}^{(i)} - \alpha
\dot{\bar{\rho}}/\bar{\rho}$,
where $\alpha$ is a first-order constant. It follows that the Fourier
coefficients of $\delta_{\rho}^{(i)}$ are gauge-invariant away from
$\bk=0$ in Fourier space.

The qualitative behaviour of the comoving density gradients can be seen
directly from their propagation equations. For scalar perturbations,
it is simplest to work directly with the equations of
motion~\eqref{eq_sec_scal_1}--\eqref{eq_sec_scal_2} for the
harmonic coefficients $\clx^{(i)}_{k}$ in the CDM frame. Eliminating the
spatial gradients of the expansion, $\clz_{a}$, we find the following
second-order equations:
\begin{eqnarray}
\ddot{\clx}^{(\gamma)}_{k} + {\tfrac{2}{3}}\theta \dot{\clx}^{(\gamma)}_{k}
+ {\tfrac{1}{3}} {\tfrac{k^{2}}{S^{2}}} \clx^{(\gamma)}_{k} &=&
{\tfrac{2}{3}}\kappa \sum_{i} \left(\rho^{(i)} + 3 p^{(i)}\right)
\clx^{(i)}_{k}
-{\tfrac{1}{3}}{\tfrac{k}{S}} \theta q^{(\gamma)}_{k} \nonumber \\
&& \mbox{} + {\tfrac{2}{3}}
{\tfrac{k^{2}}{S^{2}}} \pi^{(\gamma)}_{k} - n_{e} \sigma_{T}
{\tfrac{k}{S}} \left({\tfrac{4}{3}} v^{(b)}_{k} - q^{(\gamma)}_{k}
\right) \\
\ddot{\clx}^{(\nu)}_{k} + {\tfrac{2}{3}}\theta \dot{\clx}^{(\nu)}_{k}
+ {\tfrac{1}{3}} {\tfrac{k^{2}}{S^{2}}} \clx^{(\nu)}_{k} &=&
{\tfrac{2}{3}}\kappa \sum_{i} \left(\rho^{(i)} + 3 p^{(i)}\right)\clx^{(i)}_{k}
-{\tfrac{1}{3}}{\tfrac{k}{S}} \theta q^{(\nu)}_{k} + {\tfrac{2}{3}}
{\tfrac{k^{2}}{S^{2}}} \pi^{(\nu)}_{k}
\\
\ddot{\clx}^{(b)}_{k} + {\tfrac{2}{3}}\theta \dot{\clx}^{(b)}_{k}
+ \csound^{2} {\tfrac{k^{2}}{S^{2}}} \clx^{(b)}_{k} &=&
{\tfrac{1}{2}}\kappa \sum_{i} \left(\rho^{(i)} + 3 p^{(i)}\right)\clx^{(i)}_{k}
+ n_{e} \sigma_{T} {\tfrac{\rho^{(\gamma)}}{\rho^{(b)}}} {\tfrac{k}{S}}
\left({\tfrac{4}{3}} v^{(b)}_{k} - q^{(\gamma)}_{k}\right)
\label{eq_num01}\\
\ddot{\clx}^{(c)}_{k} + {\tfrac{2}{3}}\theta \dot{\clx}^{(c)}_{k}
&=& {\tfrac{1}{2}}\kappa \sum_{i} \left(\rho^{(i)} + 3 p^{(i)}\right)
\clx^{(i)}_{k},
\label{eq_num0}
\end{eqnarray}
where we have ignored baryon pressure except in the acoustic
term $\csound^{2} (k^{2}/S^{2}) \clx^{(b)}_{k}$, which can be significant on
small scales.
In the limiting case that the mode is well outside the Hubble radius
($\clh_{k} \gg 1$), the equations of motion for each component reduce to
the common form
\begin{equation}
\ddot{\clx}^{(i)}_{k} + {\tfrac{2}{3}}\theta \dot{\clx}^{(i)}_{k}
= \kappa {\tfrac{1}{2}} \left(1+{\tfrac{p^{(i)}}{\rho^{(i)}}}\right)
\sum_{j} \left(\rho^{(j)} + 3 p^{(j)}\right) \clx^{(j)}_{k}.
\label{eq_num1}
\end{equation}
For adiabatic initial conditions, it is clear that the adiabatic
condition, given as equation~\eqref{eq_eu_adiab}, is maintained while the
mode is outside
the Hubble radius. Solving equation~\eqref{eq_num1} for adiabatic perturbations
gives growing modes proportional to $t$ and $t^{2/3}$ during radiation
and matter domination respectively.

If a mode enters the Hubble radius prior to last scattering, the
photon/baryon fluid, which is still tightly-coupled, undergoes acoustic
oscillations. To lowest-order in the tight-coupling parameter,
$\mbox{max}(t_{c}/t_{H},t_{c}/t_{k})$, the photon
and baryon perturbations remain adiabatic, evolving according to
\begin{equation}
\ddot{\clx}^{(\gamma)}_{k} + {\tfrac{2}{3}}\theta\dot{\clx}^{(\gamma)}_{k}
+ {\tfrac{R + 3\csound^{2}}{3(1+R)}} {\tfrac{k^{2}}{S^{2}}} \clx^{(\gamma)}_{k}
= {\tfrac{2}{3}}\kappa \sum_{i} \left(\rho^{(i)} +
3 p^{(i)}\right)\clx^{(i)}_{k} - {\tfrac{4R}{9(1+R)}} {\tfrac{k}{S}}
\theta v^{(b)}_{k},
\label{eq_num2}
\end{equation}
where $R\equiv 4\rho^{(\gamma)} /3 \rho^{(b)}$. The solution of the
homogeneous equation describes acoustic oscillations in a fluid with
sound speed squared $(R+3\csound^{2})/3(1+R)$, which are damped by the
expansion of the universe. However, the oscillations are driven gravitationally
by the gradient $\sgrad_{a} (\rho + 3p)$, which gives an almost constant
amplitude oscillation in the radiation dominated era. The
Silk damping which is visible in Figure~\ref{fig1} for $k=1.0\mpc^{-1}$
at $z\simeq 10^{-3.5}$ arises from photon diffusion (which is not
described by the lowest-order tight-coupling approximation)
so is not described by equation~\eqref{eq_num2}. 
The neutrino perturbation also oscillates once inside the Hubble
radius in the radiation dominated region, while the power-law growth of the
CDM is impeded by the gravitational attraction of the oscillating dominant
component (the inhomogeneous term in eq.\ [\ref{eq_num0}]).
In the matter dominated
era, the CDM becomes the dominant component, so we again see power-law growth
of the CDM perturbation on all scales. Before last scattering, the
photons and baryons remain tightly-coupled, but the character of
the $\sgrad_{a} (\rho + 3p)$ driving term in equation~\eqref{eq_num2}
changes from
an oscillation to a power-law as the CDM becomes dominant. At last
scattering, the photons and baryons decouple. The baryons no longer feel the
pressure support provided by the photons; the Jeans' length of the
baryons is very small and the acoustic term in equation~\eqref{eq_num01} is
negligible. The $\sgrad_{a} (\rho + 3p)$ driving term attracts the baryons into
the potential wells caused principally by inhomogeneity of the CDM, so that
$\clx^{(b)}_{a}$ relaxes to $\clx^{(c)}_{a}$ as a power-law. After
last scattering, the photons and neutrinos continue to undergo driven
oscillations, which decay towards the particular integral
$\clx^{(\gamma)}_{k} = \clx^{(\nu)}_{k} = 6 \clh^{2}_{k} \clx^{(c)}_{k}$.

%%%%%%%%%%%%%%%%%%%%%%%%%%%%%%%%%%%%%%%%%%%%%%%%%%%%%%%%%%%%%%%%%%%%%%%%
\begin{figure}[t!]
\begin{center}
\epsfig{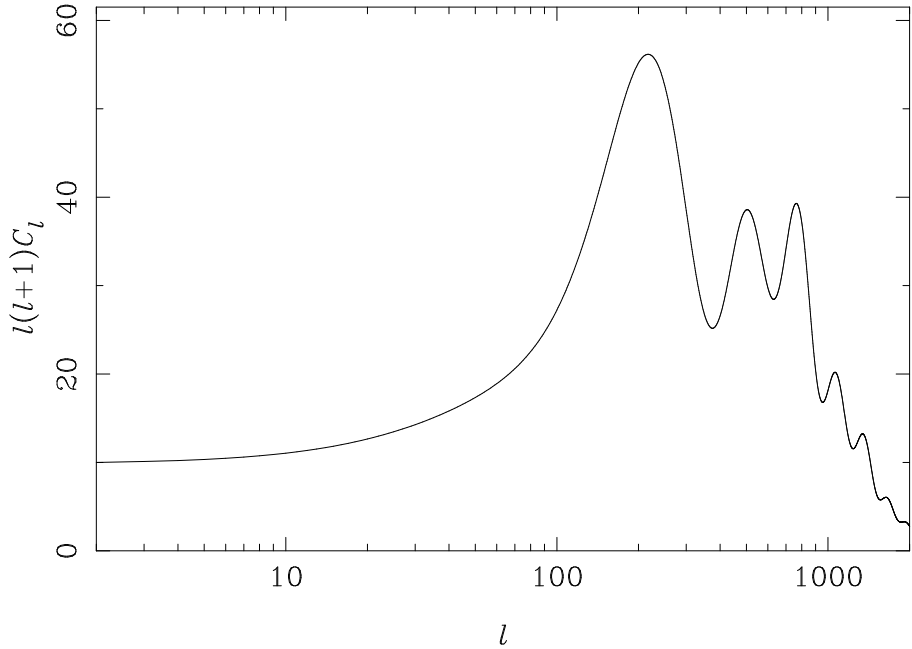}
\end{center}
\caption{The power spectrum of scalar CMB anisotropies in the
standard CDM model: $\Omega_{c}=0.95$, $\Omega_{b}=0.05$, $\Omega_{\Lambda}=0$,
$H_{0}=50\kmsmpc$,
$n_{s}=1$, with Helium fraction $0.24$. The normalisation of the vertical
scale is arbitrary.
\label{fig2}}
\end{figure}
%%%%%%%%%%%%%%%%%%%%%%%%%%%%%%%%%%%%%%%%%%%%%%%%%%%%%%%%%%%%%%%%%%%%%%%%

In Figure~\ref{fig2} we show the CMB power spectrum calculated from a
simulation in the standard CDM model. On large scales, the plateau
arises from the usual potential fluctuations $\sum_{k} \Phi_{k} \qk/3$
on the last scattering surface~\cite{sachs67}. The oscillations in
the CMB power spectrum on smaller scales (the Doppler peaks) arise from
the acoustic oscillations in the baryon/photon fluid. These oscillations
give rise to strongly scale-dependent gradients of the photon energy density
in the energy-frame, which in the approximation of instantaneous recombination
can be interpreted as temperature variations across the last scattering
surface, and a local scale-dependent distortion of the last scattering
surface relative to the energy frame. Since the last scattering surface is
well approximated by a hypersurface of constant radiation temperature
(so that recombination does occur there), it is more correct to interpret the
Doppler peaks in terms of the local variations in redshift along null
geodesics back to the last scattering surface, than in terms of temperature
variations on the last scattering surface. (There is another significant
contribution to the Doppler peaks, which is of dipole nature on the last
scattering surface, and tends to fill in the power spectrum near the
first Doppler peak; see Hu \&\ Sugiyama (1995) and Challinor \&\ Lasenby (1998)
for more details.) On the smallest
scales, the power spectrum is damped due to photon diffusion in the
photon/baryon plasma prior to recombination.

\section{Tensor Perturbations}
\label{sec_tens}

The covariant equations of Section~\ref{sec_eqs} are independent of any
non-local splitting of the perturbations into scalar, vector and tensor
modes. In the linear approximation, we have seen
that the equations describing scalar perturbations can be obtained from the
full set of equations in a straightforward manner. The same is also true for
vector and tensor perturbations. Vector perturbations are not expected to be
important today in non-seeded models since the vorticity decays in
an expanding universe~\cite{haw66}. However, most inflationary models
do predict the generation of a primordial spectrum of gravitational waves
(tensor modes) during an epoch of inflation (see Lidsey et al. (1997)
for a comprehensive review). The existence of such a background of relic
gravitational waves would have a significant effect on the CMB
anisotropy power spectrum on large scales~\cite{critt93}. For completeness,
we give the tensor multipole equations in this section. We defer
a detailed derivation of the equations, and a discussion of their relation
to the multipole equations usually employed in calculations of the effects
of tensor modes~\cite{critt93} to a future paper~\cite{chall-tens}.

In the covariant approach to cosmology, gravitational waves are characterised
by requiring that the vorticity and all gauge-invariant vectors and scalars
vanish at first-order (Dunsby et al. 1997), so that the spatial gradients of
the density and expansion vanish, as well as the acceleration and the heat
fluxes. The electric
and magnetic parts of the Weyl tensor, the shear tensor and the anisotropic
stress are constrained to be transverse:
\begin{equation}
\sgrad^{a} \cle_{ab} = 0 , \quad \sgrad^{a} \clb_{ab}=0, \quad
\sgrad^{a} \sigma_{ab} = 0, \quad \sgrad^{a} \pi_{ab} = 0.
\end{equation}
It is straightforward to show that these conditions are consistent with the
linearised propagation equations given in Section~\ref{sec_cov} (see also
Maartens (1997) for a discussion of consistency of the exact equations
for irrotational dust spacetimes). As with scalar perturbations, it is
convenient to expand harmonically the first-order covariant variables
in tensors derived from solutions of a generalised Helmholtz equation.
For tensor perturbations, we employ the tensor-valued
solutions~\cite{haw66} $\qk_{ab}=\qk_{(ab)}$
(note that we use the same symbol for the
tensor harmonics and the second-rank tensors derived from the scalar harmonic
functions to avoid cluttering formulae with additional labels), which satisfy
the zero-order relations
\begin{equation}
\qk_{ab} u^{a} = 0 ,\quad \sgrad^{a} \qk_{ab} = 0, \quad \qkdt_{ab} = 0.
\end{equation}
Expanding $\cle_{ab}$, $\clb_{ab}$, $\sigma_{ab}$
and $\pi_{ab}$ in these tensors (as in Section~\ref{sec_scal} but with
$\qk_{ab}$ replaced by the tensor harmonics, and the scalar-valued
variables $\Phi_{k}$ replaced by $\cle_{k}$), we obtain simple
propagation equations for the electric part of the Weyl tensor and the
shear:
\begin{eqnarray}
\left({\tfrac{k}{S}}\right)^{2} \left(\dot{\cle}_{k} + {\tfrac{1}{3}}
\theta\cle_{k}\right) - {\tfrac{k}{S}} \left({\tfrac{k^{2}}{S^{2}}}
+{\tfrac{3K}{S^{2}}}
- {\tfrac{1}{2}}\gamma \kappa\rho \right) \sigma_{k} &=&
-{\tfrac{1}{6}}(3\gamma-1)\kappa\rho\theta \pi_{k} + {\tfrac{1}{2}}\kappa
\rho \dot{\pi}_{k} \label{eq_tens_edt} \\
{\tfrac{k}{S}}\left(\dot{\sigma}_{k} + {\tfrac{1}{3}}\theta \sigma_{k}\right)
+ \left({\tfrac{k}{S}}\right)^{2} \cle_{k} &=& - {\tfrac{1}{2}}
\kappa \rho \pi_{k}. \label{eq_tens_sheardt}
\end{eqnarray}
Note that we have used the constraint equation~\eqref{eq_cons1} to eliminate
the magnetic part of the Weyl tensor from the propagation equation for
the electric part. Equations~\eqref{eq_tens_edt} and~\eqref{eq_tens_sheardt}
close up, with the anisotropic stress treated as a known field, to give a
second-order equation for the shear:
\begin{equation}
\ddot{\sigma}_{k} + \theta \dot{\sigma}_{k} + \left[{\tfrac{k^{2}}{S^{2}}}
+{\tfrac{2K}{S^{2}}}
- {\tfrac{1}{3}}(3\gamma-2)\kappa\rho\right]\sigma_{k} = \kappa\rho
{\tfrac{S}{k}}\left[{\tfrac{1}{3}}(3\gamma-2)\theta \pi_{k} - \dot{\pi}_{k}
\right],
\end{equation}
which generalises the homogeneous equation derived in Dunsby et al. (1997) to
include anisotropic stress.

For the photon and neutrino angular moments, $J^{(l)}_{a_{1}\ldots a_{l}}$
and $G^{(l)}_{a_{1}\ldots a_{l}}$, we expand in tensors
$\qk_{a_{1}\ldots a_{l}}$ derived from the
tensor harmonics using the same recursion relation as for scalar perturbations,
\begin{equation}
\qk_{a_{1}\ldots a_{l}} = {\tfrac{S}{k}} \left(
\sgrad_{(a_{1}} \qk_{a_{2}\ldots a_{l})} - {\tfrac{l-1}{2l-1}}
\sgrad^{b}\qk_{b(a_{1}\ldots a_{l-2}} h_{a_{l-1}a_{l})} \right),
\end{equation}
for $l\geq 2$. This procedure gives the following covariant Boltzmann
multipole equations for tensor perturbations: for the anisotropic stress
($l=2$),
\begin{eqnarray}
\dot{\pi}_{k}^{(\gamma)} + {\tfrac{1}{3}}{\tfrac{k}{S}}
\left(1-{\tfrac{6K}{k^{2}}}\right) \Jgamlink{3} - {\tfrac{8}{15}}
{\tfrac{k}{S}} \sigma_{k} &=& - {\tfrac{9}{10}} \nelec\sigma_{T}
\pi^{(\gamma)}_{k} \\
\dot{\pi}_{k}^{(\nu)} + {\tfrac{1}{3}}{\tfrac{k}{S}}
\left(1-{\tfrac{6K}{k^{2}}}\right) \Gnulink{3} - {\tfrac{8}{15}}
{\tfrac{k}{S}} \sigma_{k} &=& 0,
\end{eqnarray}
and for $l\geq 3$,
\begin{eqnarray}
\Jgamlkdt + {\tfrac{k}{S}} \left\{ {\tfrac{(l+3)(l-1)}{(l+1)(2l+1)}}
\left[1-\left((l+1)^{2}-3\right){\tfrac{K}{k^{2}}} \right]
\Jgamlink{l+1} - {\tfrac{l}{2l+1}}\Jgamlink{l-1} \right\}
&=& - \nelec \sigma_{T} \Jgamlink{l} \\
\Gnulkdt + {\tfrac{k}{S}} \left\{ {\tfrac{(l+3)(l-1)}{(l+1)(2l+1)}}
\left[1-\left((l+1)^{2}-3\right){\tfrac{K}{k^{2}}} \right]
\Gnulink{l+1} - {\tfrac{l}{2l+1}}\Gnulink{l-1} \right\}
&=& 0.
\end{eqnarray}
As we stressed earlier, by performing the covariant angular expansion before
harmonically expanding the moments, the necessary angular dependence of the
moments appears automatically, whereas in the metric-based approach, this
(azimuthal) dependence must be put in by hand, and is different for the
two polarisations of gravitational waves. It should be noted that the
moment equations given here are not the same as those satisfied by
the $\tilde{\Delta}_{Il}^{i}$ variables, where $i=+$, $\times$ labels
the polarisation of the gravitational wave and $I$ denotes the intensity
Stokes parameter (we follow the notation employed in Kosowsky (1996)), that
are usually employed in metric-based calculations of the effects of tensor
modes. This is because the covariant angular expansion gives rise to a more
natural set of variables, the $\Jgamlink{l}$, which are related to the
temperature anisotropy in a simpler manner than the
$\tilde{\Delta}_{Il}^{i}$. In particular, the $l$-th multipole $C_{l}$
of the anisotropy power spectrum depends only on $\Jgamlink{l}$, whereas
$C_{l}$ depends on the $(l-2)$, $l$ and $(l+2)$-th moments,
$\tilde{\Delta}_{Il}^{i}$, thus obscuring the physical interpretation of these
variables. The relation between the two sets of variables will be discussed
further in Challinor (1998).

The first-order propagation equations for the shear and electric part of the
Weyl tensor, along with the Boltzmann multipole equations for the photons and
neutrinos give a closed set of equations that can be solved to calculate the
temperature anisotropy for given initial conditions. The numerical solution
of these equations will be considered elsewhere~\cite{chall-tens}.

\section{Conclusion}
\label{sec_conc}

We have shown how the full kinetic-theory calculation of the evolution of
CMB anisotropies and density inhomogeneities can be performed in the
covariant approach to cosmology (Ehlers 1993; Ellis 1971), using the
gauge-invariant perturbation theory of Ellis \&\ Bruni (1989).
Adopting covariantly-defined, gauge-invariant variables throughout
ensured that our discussion avoided the gauge ambiguities that appear in
certain gauges, and that all variables had a clear, physical interpretation.
We presented a unified set of equations describing the evolution of photon
and neutrino anisotropies and cosmological perturbations in the CDM model,
which were independent of a decomposition into scalar, vector or tensor
modes and the associated harmonic analysis. Although we only
considered the case of linear perturbations around an FRW universe here,
it is straightforward to extend the approach to include non-linear
effects (Maartens et al. 1998), which should allow a physically transparent
discussion of second-order effects on the CMB.
Indeed, the ease with which one can
write down the exact equations for the physically relevant variables
is one of the major strengths of the covariant approach.

The linear equations describing scalar modes and tensor modes were obtained
from the full set of equations in a straightforward and unified manner,
highlighting the advantage of having the full equations (independent
of the decomposition into scalar, vector and tensor modes) available.
For the scalar case, the Boltzmann multipole equations for the moments
of the distribution functions obtained here were equivalent to those
usually seen in the literature. However, for tensor modes, the covariant
approach led naturally to a set of moment variables that more
conveniently describe the temperature anisotropy than those usually employed.
For scalar modes, we discussed the solution of the perturbation equations
in detail, including the integral solution of the Boltzmann multipole
equations and the relation between the timelike integrations performed in
the multipole approach to calculating CMB anisotropies, and the lightlike
integrations of the line of sight approach. The numerical solution of the
scalar equations in a $K=0$, almost-FRW, CDM universe were also discussed.
Our numerical results provide independent confirmation of those of
other groups, (see, for example, Ma \&\ Bertschinger (1995) and Seljak \&\
Zaldarriaga (1996)), who have obtained their
results by employing non-covariant methods in specific gauges. Typically,
these methods require one to keep careful track of all residual gauge-freedom,
both to enable identification of any gauge-mode solutions, and to ensure that
the final results quoted are gauge-invariant (and hence observable).
Fortunately, the isotropy of the photon distribution function in
an exact FRW universe ensures that the CMB power spectrum, as calculated from
the gauge-dependent perturbation to the distribution function, is
gauge-invariant for $l\geq 1$.

We hope to have shown the ease with which the covariant approach to
cosmology can be applied to the problem of calculating CMB anisotropies.
The covariant and gauge-invariant method discussed here frees one from the
gauge problems that have caused confusion in the past, and focuses
attention on the physically relevant variables in the problem and the
underlying physics. Future work in this area will include the discussion
of non-linear effects (Maartens at al. 1998), the inclusion of polarisation,
and the effects of hot dark matter, all of which can be expected to
bring the same advantages of physical clarity and transparency that we hope
to have demonstrated here.

\acknowledgements

The development of the COSMICS package was supported by the NSF under grant
AST-9318185. The authors wish to thank Roy Maartens for useful comments on
an earlier version of this paper.

\end{document}